\documentclass[reprint,amsmath,amssymb,aps,prd,]{revtex4-1}
\usepackage{graphicx}
\usepackage{epsfig,graphics,subfigure,psfrag,amsmath,amssymb}
\usepackage{tabularx}
\usepackage{adjustbox}  
\usepackage{dcolumn}
\usepackage{bm}
\usepackage{overpic}
\usepackage{xspace}
\usepackage{rotating}
\usepackage{color}
\usepackage{multirow}
\usepackage{colortbl}
\usepackage{epstopdf}
\usepackage{float}
\usepackage{lineno}
\usepackage{amsmath}
\definecolor{aa}{RGB}{0,0,255}
\usepackage[bookmarksnumbered=true,colorlinks,urlcolor=aa,linkcolor=blue,anchorcolor=aa,citecolor=aa]{hyperref}

\hfuzz=\maxdimen
\vfuzz=\maxdimen
\makeatletter
\let\HyperrefWarn\@gobble
\makeatother

\newcommand{\bfx}{\mathbf{x}}

\newcommand{\BRPsipGamchicZero    }{9.79 \pm 0.20}
\newcommand{\BRPsipGamchicOne     }{9.75 \pm 0.24}
\newcommand{\BRPsipGamchicTwo     }{9.52 \pm 0.20}
\newcommand{\BRLambdaPpi          }{63.90 \pm 0.50}
\newcommand{\ProdBFChicZero      }{3.585 \pm 0.047 \pm 0.080}
\newcommand{\ProdBFChicOne       }{1.152 \pm 0.025 \pm 0.030}
\newcommand{\ProdBFChicTwo       }{1.622 \pm 0.033 \pm 0.042}

\newcommand{\Nobs                }{12831\pm114} 
\newcommand{\Ndata               }{12880} 
\newcommand{\Nbg                 }{49} 
\newcommand{\Efficiency          }{17.64\pm 0.05} 
\newcommand{\BFChicZero          }{3.662 \pm 0.048 \pm 0.111}
\newcommand{\BFChicOne           }{1.182 \pm 0.026 \pm 0.042}
\newcommand{\BFChicTwo           }{1.704 \pm 0.035 \pm 0.057}
\newcommand{\rOne                }{1.037     \pm 0.047    \pm 0.029   }
\newcommand{\rTwo                }{1.48      \pm 0.13     \pm 0.07    }
\newcommand{\rThree              }{1.85      \pm 0.15     \pm 0.10    }
\newcommand{\DeltaPhiROne        }{0.042     \pm 0.107    \pm 0.032   }
\newcommand{\DeltaPhiRTwo        }{0.37      \pm 0.12     \pm 0.17    }
\newcommand{\DeltaPhiRThree      }{0.13      \pm 0.12     \pm 0.05    }
\newcommand{\RChicTwo            }{0.575     \pm 0.048    \pm 0.018   }
\newcommand{\DeltaPhiChicTwo     }{0.37      \pm 0.15     \pm 0.05    }

\newcommand{\FFSJChicZero          }{0.5459    \pm 0.0052   \pm 0.0076  }
\newcommand{\FFSJChicOne           }{0.1755    \pm 0.0036   \pm 0.0034  }
\newcommand{\FFSJChicTwo           }{0.2470    \pm 0.0046   \pm 0.0047  }
\newcommand{\FFNRTwoPlus         }{0.0265    \pm 0.0040   \pm 0.0098  }
\newcommand{\AlphaChicOne        }{-0.301    \pm 0.042    \pm 0.026   }
\newcommand{\AlphaChicTwo        }{-0.211    \pm 0.100    \pm 0.050   }
\newcommand{\BetaChicTwo         }{-0.039    \pm 0.089    \pm 0.033   }
\newcommand{\GammaChicZero       }{12.31   \pm 0.26  \pm 0.12 }
\newcommand{\EventschicZero    }{7437 \pm 64}
\newcommand{\EventschicOne     }{2278 \pm 45}
\newcommand{\EventschicTwo     }{2787 \pm 56}
\newcommand{\EventsPsipA    }{107.7 \pm 0.6}
\newcommand{\EventsPsipB    }{345.4 \pm 2.6}
\newcommand{\EventsPsipC    }{2259.3 \pm 11.1}

\newcommand{\LumiPsip           }{2712.4 \pm 14.3} 
\newcommand{\LambdaMassWindow    }{[1.108,1.123]}
\newcommand{\EMCBarrelRegion    }{|\cos\theta|<0.8}
\newcommand{\EMCEndCapRegion    }{0.86<|\cos\theta|<0.92}
\newcommand{\ChisqCut    }{40} 
\newcommand{\SigmaMassWindow    }{[1.183,1.216]} 
\newcommand{\ASigmaMassWindow    }{[1.183,1.219]}
\newcommand{\ChicjMassWindow    }{[3.20,3.58]}
\newcommand{\BackgroundRate    }{0.4\%}
\newcommand{\PDGBFChicZero    }{3.61 \pm 0.16}
\newcommand{\PDGBFChicOne    }{1.27 \pm 0.09}
\newcommand{\PDGBFChicTwo    }{1.86 \pm 0.16}

\begin{document}

\title{\boldmath Helicity amplitude  and branching fraction measurement of $\chi_{cJ} \rightarrow \Lambda\bar{\Lambda} $ }

\newcommand{\BESIIIorcid}[1]{\href{https://orcid.org/#1}{\hspace*{0.1em}\raisebox{-0.45ex}{\includegraphics[width=1em]{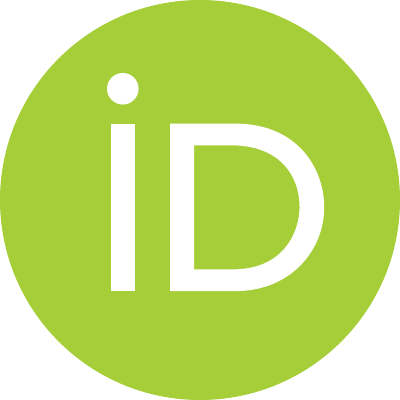}}}}

\author{
  \begin{small}
    \begin{center}
M.~Ablikim$^{1}$\BESIIIorcid{0000-0002-3935-619X},
M.~N.~Achasov$^{4,b}$\BESIIIorcid{0000-0002-9400-8622},
P.~Adlarson$^{77}$\BESIIIorcid{0000-0001-6280-3851},
X.~C.~Ai$^{82}$\BESIIIorcid{0000-0003-3856-2415},
R.~Aliberti$^{36}$\BESIIIorcid{0000-0003-3500-4012},
A.~Amoroso$^{76A,76C}$\BESIIIorcid{0000-0002-3095-8610},
Q.~An$^{59,73,\dagger}$,
Y.~Bai$^{58}$\BESIIIorcid{0000-0001-6593-5665},
O.~Bakina$^{37}$\BESIIIorcid{0009-0005-0719-7461},
Y.~Ban$^{47,g}$\BESIIIorcid{0000-0002-1912-0374}
% ...
\\
\vspace{0.2cm}
(BESIII Collaboration)\\
\vspace{0.2cm} {\it
$^{1}$ Institute of High Energy Physics, Beijing 100049, People's Republic of China\\
% ...
$^{4}$ Budker Institute of Nuclear Physics SB RAS (BINP), Novosibirsk 630090, Russia\\
% ...
$^{36}$ Johannes Gutenberg University of Mainz, Johann-Joachim-Becher-Weg 45, D-55099 Mainz, Germany\\
$^{37}$ Joint Institute for Nuclear Research, 141980 Dubna, Moscow region, Russia\\
% ...
$^{47}$ Peking University, Beijing 100871, People's Republic of China\\
% ...
$^{58}$ Southeast University, Nanjing 211100, People's Republic of China\\
$^{59}$ State Key Laboratory of Particle Detection and Electronics, Beijing 100049, Hefei 230026, People's Republic of China\\
% ...
$^{73}$ University of Science and Technology of China, Hefei 230026, People's Republic of China\\
% ...
$^{76}$ University of Turin and INFN, (A)University of Turin, I-10125, Turin, Italy; (B)University of Eastern Piedmont, I-15121, Alessandria, Italy; (C)INFN, I-10125, Turin, Italy\\
$^{77}$ Uppsala University, Box 516, SE-75120 Uppsala, Sweden\\
% ...
$^{82}$ Zhengzhou University, Zhengzhou 450001, People's Republic of China\\

\vspace{0.2cm}
$^{\dagger}$ Deceased\\
% ...
$^{b}$ Also at the Novosibirsk State University, Novosibirsk, 630090, Russia\\
% ...
$^{g}$ Also at State Key Laboratory of Nuclear Physics and Technology, Peking University, Beijing 100871, People's Republic of China\\
% ...

}

\end{center}
    \vspace{0.4cm}
\end{small}}
\affiliation{}


\begin{abstract}
Utilizing $\LumiPsip$ million $\psi(3686)$ events accumulated by the
BESIII experiment, we perform a partial wave analysis of
$\psi(3686)\rightarrow\gamma\chi_{cJ}\rightarrow\gamma\Lambda\bar{\Lambda}$
decay ($J=0,1,2$). The ratio of the helicity amplitudes with same (++)
and opposite (+-) helicity for
$\chi_{c2}\rightarrow\Lambda\bar{\Lambda}$ decay is determined for the
first time to be $R_{\chi_{c2}}=\RChicTwo$, with a relative phase
angle $\Delta\Phi_{\chi_{c2}} = \DeltaPhiChicTwo$~rad.  The parameters of
the angular distribution of $\chi_{c2}$ are determined to be
$\alpha_{\chi_{c2}} = \AlphaChicTwo$ and $\beta_{\chi_{c2}} =
\BetaChicTwo$, based on the distribution $dN / d\cos\theta = 1 +
\alpha_{\chi_{c2}} \cos^2\theta + \beta_{\chi_{c2}} \cos^4\theta$. The
width of $\chi_{c0}$ is determined to be $\GammaChicZero$~MeV. Additionally, the branching fractions for $\chi_{cJ} \rightarrow
\Lambda\bar{\Lambda}$ are measured to be $(\BFChicZero) \times
10^{-4}$, $(\BFChicOne) \times 10^{-4}$, and $(\BFChicTwo) \times
10^{-4}$ for $\chi_{c0}$, $\chi_{c1}$ and $\chi_{c2}$, respectively,
where the first uncertainty is statistical and the second systematic.
\end{abstract}

\maketitle

\renewcommand{\arraystretch}{1.3}

\section{INTRODUCTION}

The $P$-wave triplet charmonium states \mbox{$\chi_{cJ} (J=0,1,2)$},
composed of a charm-anticharm quark pair, lying in a region between
perturbative and non-perturbative Quantum Chromodynamics (QCD), have
been known for more than forty years, yet their decay mechanisms
remain unclear. To understand them, especially their structure and
non-perturbative QCD properties, it is crucial to experimentally
investigate these charmonium states to a greater extent. Their baryonic
decay branching fractions, in particular, have been extensively
studied in various experiments, achieving roughly a few percent
precision
~\cite{PhysRevD.87.032007,PhysRevD.97.052011,PhysRevD.101.092002,PhysRevD.83.112009,BESIII:2023olq}.

In the asymptotic energy region of massless perturbative QCD, the
vector coupling of gluons and quarks conserves the helicity of the
quarks, which leads to the helicity selection rule (HSR)
\cite{Brodsky:1981kj}.  The helicity-violating processes, such as
$\eta_{c},\chi_{c0}\rightarrow B \bar{B}$, where $B$ denotes baryon,
are forbidden by HSR.  However, experimental measurements have shown
that the branching fractions of $\chi_{c0}$ decaying to baryonic final
states are comparable to those of $\chi_{c1}$ and $\chi_{c2}$ decays
\cite{pdg}.  Those observations suggest that non-perturbative effects
may play a significant role in $\chi_{cJ}$ decays.

Some theories have explored this phenomenon from the perspective of
charmonium structure, suggesting that charmonium can exhibit
HSR-violating decays through mixing with glueballs
\cite{Anselmino:1991es} or the color octet mechanism
\cite{Wong:1999hc}. HSR-violation due to non-vanishing quark mass
has also been investigated. It was found that by assigning quarks a
constituent mass rather than a current mass, a comparable branching
fraction of $\chi_{c0}\rightarrow p \bar{p}$ can be obtained as
for the non-forbidden processes $\chi_{c1,2}\rightarrow p \bar{p}$
\cite{Anselmino:1992jd}. Other mechanisms, including quark pair
creation~\cite{Ping:2004sh}, quark-diquark model
\cite{Anselmino:1991mc}, and the long-distance contribution via
charmed hadron loops~\cite{Liu:2010um}, have also been
investigated. More recently, with the availability of precise
experimental data, a systematic investigation based on SU(3) flavor
symmetry has provided a unified framework to understand these
HSR-violating processes~\cite{Lan:2024lvs}. This approach successfully
describes both $\chi_{cJ} \to B_8\bar{B}_8$ and $B_{10}\bar{B}_{10}$
decays within 2 standard deviations after including SU(3) breaking
effects. Here $B_8$ and $B_{10}$ represent light octet and decuplet
baryons, respectively.

Precise measurement results of $\chi_{cJ}$ decays can be used to test
theoretically proposed decay mechanisms and provide a deeper insight
into the structure and properties of $\chi_{cJ}$ states. The BESIII
detector has accumulated ($\LumiPsip$) million $\psi(3686)$
events~\cite{BESIII:2024lks}, which offers a good opportunity to study
$\chi_{cJ}$ decays with high precision. In this paper, we present
a partial wave analysis (PWA) for the process $\psi(3686) \rightarrow
\gamma\Lambda\bar{\Lambda} \rightarrow \gamma p\pi^-\bar{p}\pi^+$,
providing a precise measurement of the helicity amplitude ratio of
$\chi_{c2}\rightarrow\Lambda\bar{\Lambda}$, the branching fractions of
$\chi_{cJ}\rightarrow\Lambda\bar{\Lambda}$ ($J=0,1,2$), and the width
of $\chi_{c0}$.

\section{BESIII DETECTOR AND MONTE CARLO SIMULATION}
\label{MCsimulation}

The BESIII detector~\cite{Ablikim:2009aa} records symmetric $e^+e^-$
collisions provided by the BEPCII storage
ring~\cite{Yu:IPAC2016-TUYA01} in the center-of-mass energy range from
1.84 to 4.95~GeV, with a peak luminosity of $1.1 \times
10^{33}$~cm$^{-2}$s$^{-1}$ achieved at $\sqrt{s} =
3.773$~GeV. BESIII has collected large data samples in this
energy region~\cite{Ablikim:2019hff,EcmsMea,EventFilter}. The
cylindrical core of the BESIII detector covers 93\% of the full solid
angle and consists of a helium-based
 multilayer drift chamber~(MDC), a plastic scintillator time-of-flight
system~(TOF), and a CsI(Tl) electromagnetic calorimeter~(EMC),
which are all enclosed in a superconducting solenoidal magnet
providing a 1.0~T  magnetic field. The solenoid is supported by an
octagonal flux-return yoke with resistive plate counter muon
identification modules interleaved with steel. 
The charged-particle momentum resolution at $1~{\rm GeV}/c$ is
0.5\%, and the 
${\rm d}E/{\rm d}x$
resolution is 6\% for electrons
from Bhabha scattering. The EMC measures photon energies with a
resolution of 2.5\%~(5\%) at 1~GeV in the barrel (end cap)
region. The time resolution in the TOF barrel region is 68~ps, while
that in the end cap region was 110~ps.  The end cap TOF
system was upgraded in 2015 using multigap resistive plate chamber
technology, providing a time resolution of
60~ps,
which benefits 83\% of the data used in this analysis~\cite{etof1,etof2,etof3}.

Monte Carlo (MC) samples are generated using the {\textsc
geant4}-based~\cite{geant4} BESIII Offline Software
System~\cite{Ablikim:2009aa}, which includes the geometric description
of the BESIII detector and the detector response, to determine
detection efficiencies and to estimate backgrounds. The simulation
models the beam energy spread and initial state radiation (ISR) in the
$e^+e^-$ annihilations with the generator {\textsc
KKMC}~\cite{ref:kkmc}. The inclusive MC sample includes the production
of the $\psi(3686)$ resonance, the ISR production of the $J/\psi$, and
the continuum processes incorporated in {\textsc
KKMC}~\cite{ref:kkmc}. All particle decays are modeled with {\textsc
evtgen}~\cite{ref:evtgen} using branching fractions either taken from
the Particle Data Group (PDG)~\cite{pdg}, when available, or otherwise
estimated with {\textsc lundcharm}~\cite{ref:lundcharm}. Three
$\psi(3686)$ data samples collected in 2009, 2012, and 2021 with
$(\EventsPsipA)$, $(\EventsPsipB)$, and $(\EventsPsipC)$ million
events~\cite{BESIII:2024lks}, respectively, and their corresponding inclusive MC
events, are utilized in the analysis. Also 10 million signal MC events
are generated using a PWA model for the decay
$\psi(3686)\to\gamma\chi_{cJ}\rightarrow\gamma\Lambda\bar{\Lambda}$.
Final state radiation from charged final state particles is
incorporated using {\textsc photos}~\cite{photos}.

\section{EVENT SELECTION} \label{sec_sel}

Events from $\psi(3686)\to\gamma\chi_{cJ}$,
$\chi_{cJ}\to\Lambda\bar{\Lambda}$ decay are identified by selecting
events containing one photon and a pair of $\Lambda$-$\bar{\Lambda}$
particles. The $\Lambda(\bar{\Lambda})$ particle is reconstructed by
its dominant decay $\Lambda \rightarrow p\pi^-$ ($\bar{\Lambda}
\rightarrow \bar{p}\pi^+$). Charged tracks detected in MDC are
required to have \mbox{$|\cos(\theta)|<0.93$} where $\theta$ is the
polar angle of the track with respect to the beam axis. The distance of closest approach between track and the
interaction point is required to be less than 20 cm along the beam axis. 
This requirement ensures that the tracks originate from the $e^+e^-$ 
annihilation point and suppresses contamination from cosmic ray tracks. 
Particle identification (PID) for charged tracks combines
measurements of the energy deposited in the MDC~(d$E$/d$x$) and the
flight time in the TOF to form likelihoods $\mathcal{L}(h)~(h=p,\pi)$
\cite{Asner:2008nq} for each hadron $h$ hypothesis. Tracks are
identified as protons when the proton hypothesis likelihood is larger
than that of the pion and larger than zero
($\mathcal{L}(p)>\mathcal{L}(\pi)$ and $\mathcal{L}(p)>0$), otherwise
they will be regarded as pions. The PID efficiency is nearly 100\% for high momentum protons \cite{BESIII:2023ooh}. 
At least one candidate for each of $p$, $\bar{p}$, $\pi^-$, and $\pi^+$ is required.

\begin{figure}[h]
  \centering
\subfigure{\includegraphics[width=0.235\textwidth]{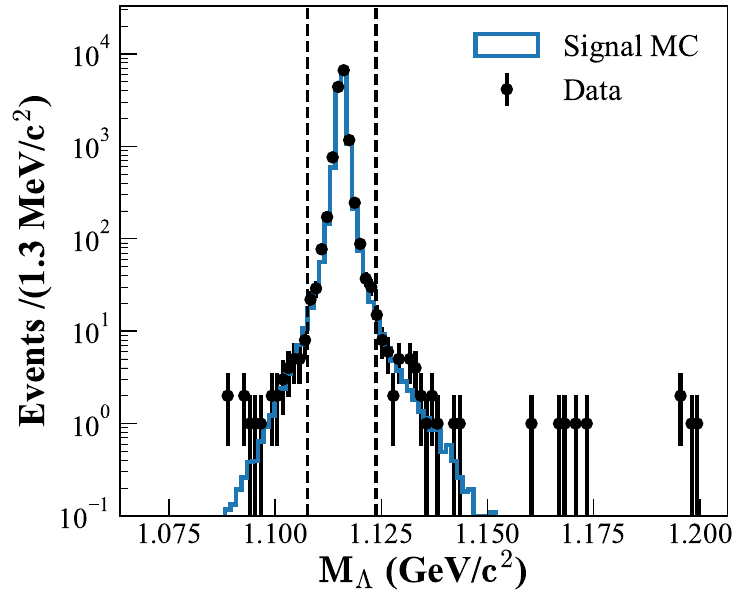}}
  \subfigure{\includegraphics[width=0.235\textwidth]{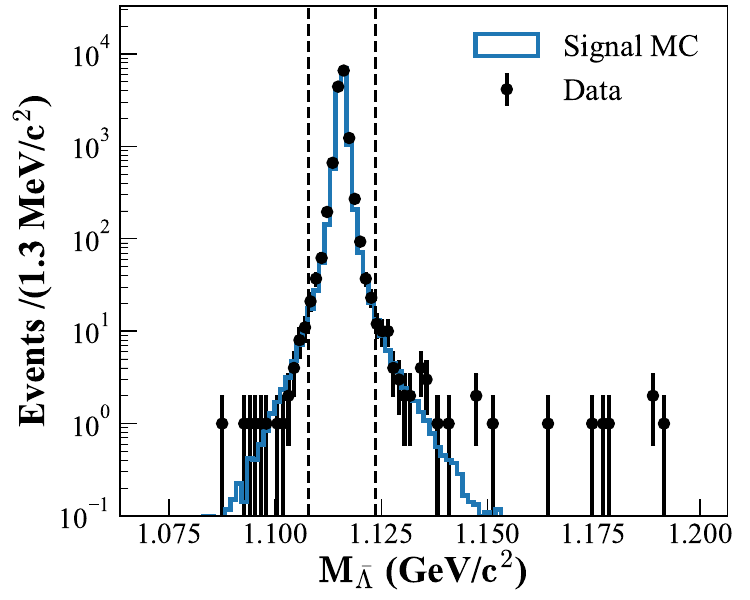}}
  \caption{Invariant  mass distribution of $\Lambda$ and $\bar{\Lambda}$ reconstructed by $p\pi^-$ and $\bar{p}\pi^+$ respectively. The two vertical dashed lines represent the mass window \LambdaMassWindow~GeV/$c^2$. }
  \label{fig:invariant_mass}
\end{figure}

The $\Lambda$ and $\bar{\Lambda}$ candidates are identified by examining all possible $(p\pi^-)(\bar{p}\pi^+)$ combinations and selecting the combination with the smallest $\chi^2=(M_{p \pi^-} - m_{\Lambda})^2 + (M_{\bar{p}\pi^+} - m_{\bar{\Lambda}})^2$, where $M_{p \pi^-}$ and $M_{\bar{p}\pi^+}$ are the invariant masses of $(p\pi^-)$ and $(\bar{p}\pi^+)$ systems respectively, and $m_{\Lambda}$ and $m_{\bar{\Lambda}}$ are the nominal masses of $\Lambda$ and $\bar{\Lambda}$ respectively~\cite{ParticleDataGroup:2024cfk}.
To improve the momentum resolution, vertex constrained fits are applied to the $\Lambda$ and $\bar{\Lambda}$ candidates, followed by secondary vertex constrained fits~\cite{Xu:2009zzg}. 
Given the long lifetime of $\Lambda(\bar{\Lambda})$ particles, the decay length divided by its uncertainty is required to be greater than 2.0 to suppress non-$\Lambda(\bar{\Lambda})$ backgrounds. 
Furthermore, a mass window of \LambdaMassWindow~GeV/$c^2$ is applied to both $\Lambda$ and $\bar{\Lambda}$ candidates, as shown in Fig.~\ref{fig:invariant_mass}.

The radiative photon is identified using showers in EMC.  The
deposited energy of each shower must be greater than 25~MeV in the
barrel region \mbox{($\EMCBarrelRegion$)} and greater than 50~MeV in
the end cap region \mbox{($\EMCEndCapRegion$)}. The difference
between the EMC time and the event start time of each shower is
required to be within [0,700]~ns to suppress the electronic noise and
energy deposition unrelated to the signal event. A four-constraint
(4C) kinematic fit, where the total four-momentum is constrained to
match the initial four-momentum of the $\psi(3686)$, is performed on
the $\Lambda\bar{\Lambda}$ with each of the showers. The one with the
minimum $\chi^2_{4C}$ is selected as the radiative photon candidate,
and the $\chi^2_{4C}$ is required to be less than \ChisqCut to suppress
background.
 
The main backgrounds remaining arise from the processes
$\psi(3686)\rightarrow\Sigma^0\bar{\Sigma}^0$ and $\psi(3686)
\rightarrow \Lambda\bar{\Sigma}^0 + c.c.$. By requiring the invariant mass of
$\gamma\Lambda(\gamma\bar{\Lambda})$ to be outside of the
$\Sigma^0$($\bar{\Sigma}^0$) mass window
$\SigmaMassWindow(\ASigmaMassWindow)$~GeV/$c^2$, 93\% $\psi(3686)\rightarrow\Sigma^0\bar{\Sigma}^0$ backgrounds and 99\% $\psi(3686)\rightarrow \Lambda\bar{\Sigma}^0 + c.c.$ backgrounds are removed. The mass windows for
$\Sigma^0$ and $\bar{\Sigma}^0$ differ slightly due to different mass
resolutions.

\begin{figure*}[htbp]
\subfigure{\includegraphics[width=0.315\textwidth]{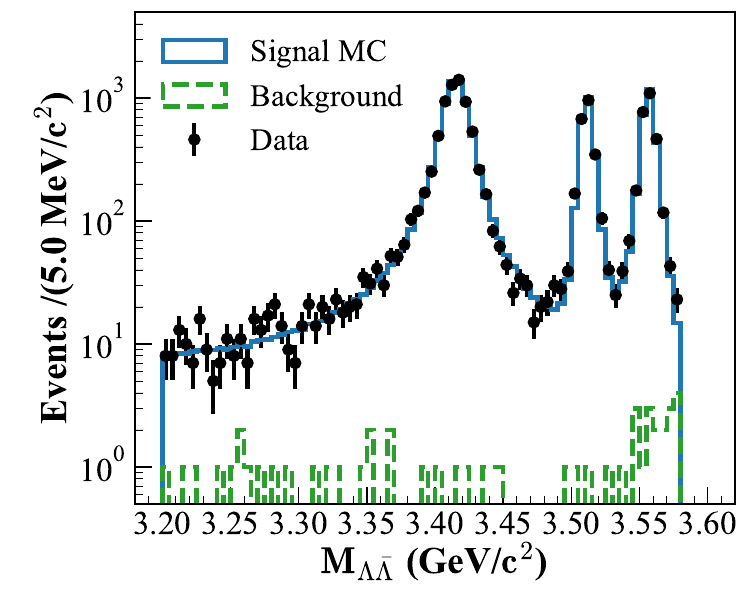}}
  \subfigure{\includegraphics[width=0.315\textwidth]{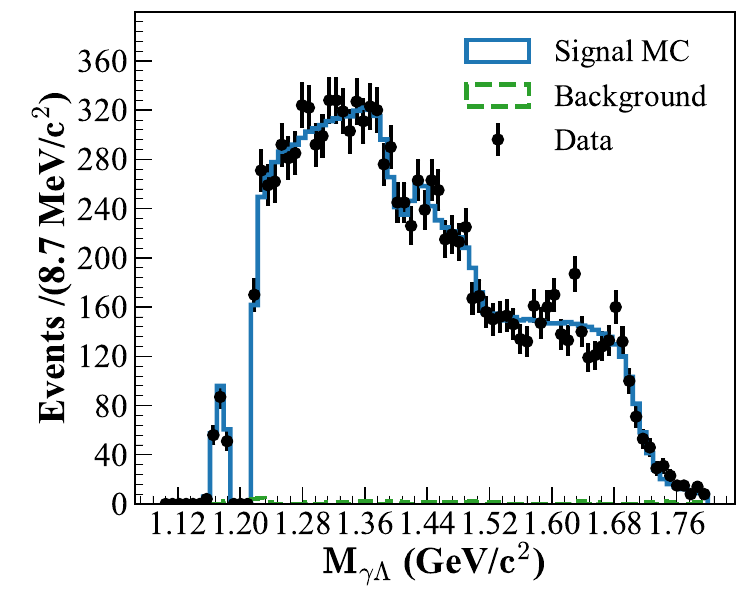}}
  \subfigure{\includegraphics[width=0.315\textwidth]{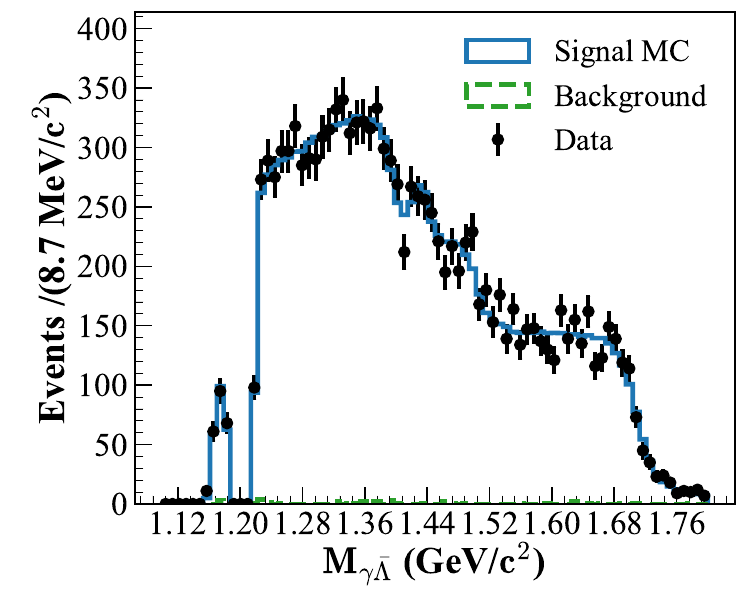}}
  \caption{Invariant mass distributions of $\Lambda\bar{\Lambda}$, $\gamma\Lambda$ and
$\gamma\bar{\Lambda}$ for the selected events. The
  signal MC is normalized to the data by area normalization, excluding the
  background component. The three peaks in the left panel correspond to the $\chi_{c0}$, $\chi_{c1}$, 
and $\chi_{c2}$ signals, respectively. The dip around 1.2 GeV/$c^2$ in the middle and right panels 
is caused by the $\Sigma^0$ and $\bar{\Sigma}^0$ veto requirements.} 
  \label{fig:invariant_mass_2}
\end{figure*}

Figure \ref{fig:invariant_mass_2} shows the invariant mass
distribution of $\Lambda\bar{\Lambda}$, $\gamma\Lambda$ and
$\gamma\bar{\Lambda}$ after using the above selection criteria. The
signal range for $\chi_{cJ}$ is $M_{\Lambda\bar{\Lambda}}\in$
\ChicjMassWindow~GeV/$c^2$, and the number of selected data candidates is
$\Ndata$. The number of background events is estimated to be $\Nbg$
from the inclusive MC sample, corresponding to a background rate
of \BackgroundRate.

A PWA fitting is performed on the selected events. The amplitude
formula and the fitting results are presented in Sec.~\ref{amplitudeformula} and~\ref{sec_fit}, respectively.

\section{AMPLITUDE FORMULA}
\label{amplitudeformula}


We construct the amplitude using the helicity formalism~\cite{Chung:1971ri,Richman:1984gh}, which defines the amplitude for a decay process using helicity eigenstates. Helicity is the projection of a particle's spin onto the direction of its momentum, commonly denoted by $\lambda$.  The decay amplitude for each sequential decay can be characterized by
the helicity angles $\Omega_i=(\theta_i,\phi_i)$ and helicities of particles, defined as follows:

\begin{itemize}

\item In the
$e^+e^-\to\psi(3686)\to\gamma(\lambda_1)\chi_{cJ}(\lambda_2)$
  ($\lambda_1 = -1, 1; -J \le \lambda_2 \le  J$) process,
the helicity system is established with the $z$-axis aligned with the
$e^+$ particle's direction. The angles
$(\theta_0,\phi_0)$ describe the orientation of the $\chi_{cJ}$
particle in the center-of-mass frame of the $e^+e^-$ collision. 
The helicity amplitude of this decay is denoted by $A_{\lambda_1,\lambda_2}^{(J)}$.
This electromagnetic radiation decay conserves parity and angular momentum,
resulting in the following relations:
$A^{(0)/(2)}_{\lambda_1,\lambda_2}=A^{(0)/(2)}_{-\lambda_1,-\lambda_2}$,
$A^{(1)}_{\lambda_1,\lambda_2}=-A^{(1)}_{-\lambda_1,-\lambda_2}$, and
$A^{(J)}_{\lambda_1,\lambda_2}=0$ if $|\lambda_1-\lambda_2|>J$.

\item In the $\chi_{cJ}\to\Lambda(\lambda_3)\bar\Lambda(\lambda_4)$
  ($\lambda_3,\lambda_4 = -1/2, 1/2$)
decay, the azimuthal angle $\phi_1$ denotes the separation between the
decay and production planes of $\chi_{cJ}$. $\theta_1$ is then the
angle between the momentum of $\Lambda$ in the rest frame of
$\chi_{cJ}$ and the momentum of $\chi_{cJ}$ in the center-of-mass frame. 
The helicity amplitude of this decay is denoted by $B_{\lambda_3,\lambda_4}^{(J)}$.
Parity conservation leads
to the following relations:
$B^{(0)/(2)}_{\lambda_3,\lambda_4}=B^{(0)/(2)}_{-\lambda_3,-\lambda_4}$
and $B^{(1)}_{\lambda_3,\lambda_4}=-B^{(1)}_{-\lambda_3,-\lambda_4}$,
while CP conservation implies $B^{(1)}_{\lambda_3,\lambda_4}=0$ if
$\lambda_3=\lambda_4$ for $\chi_{c1}$ decays.

\item For the $\Lambda\to p(\lambda_5)\pi^-$ ($\lambda_5 = -1/2, 1/2$) decay, $\phi_2$ is the
angle between the production and decay planes of $\Lambda$, while
$\theta_2$ is the angle between the proton's momentum in the rest
frame of $\Lambda$ and the momentum of $\Lambda$ in the rest frame of
$\chi_{cJ}$. The helicity amplitude of this decay is denoted by $F_{\lambda_5}$, 
which is related to the $\Lambda$ decay asymmetry parameter by
$\alpha_{\Lambda}=\frac{|F_{+}|^2-|F_{-}|^2}{|F_{+}|^2+|F_{-}|^2}$
\cite{Chen:2020pia_Pchicj}, with $F_{\pm} \equiv F_{\pm 1/2}$.
In the fit, $\alpha_{\Lambda}$ is fixed to the value measured by BESIII~\cite{BESIII:2022qax}.

\item In the $\bar{\Lambda}\to \bar p(\lambda_6)\pi^+$ ($\lambda_6 = -1/2, 1/2$) decay, $\phi_3$
represents the angle between the production and decay planes of
$\bar{\Lambda}$, and $\theta_3$ is the angle between the anti-proton's
momentum in the rest frame of $\bar{\Lambda}$ and the momentum of
$\bar{\Lambda}$ in the rest frame of $\chi_{cJ}$. The helicity amplitude of this decay is denoted by 
$G_{\lambda_6}$, which is related to the $\bar\Lambda$ decay asymmetry parameter by
$\alpha_{\bar\Lambda}=\frac{|G_{+}|^2-|G_{-}|^2}{|G_{+}|^2+|G_{-}|^2}$
\cite{Chen:2020pia_Pchicj}, with $G_{\pm} \equiv G_{\pm 1/2}$.
The $\alpha_{\bar\Lambda}$ is constrained to equal $\alpha_{\Lambda}$ by CP symmetry.
\end{itemize} \ \

\begin{figure}[H]
\centering
\includegraphics[clip,width=\columnwidth]{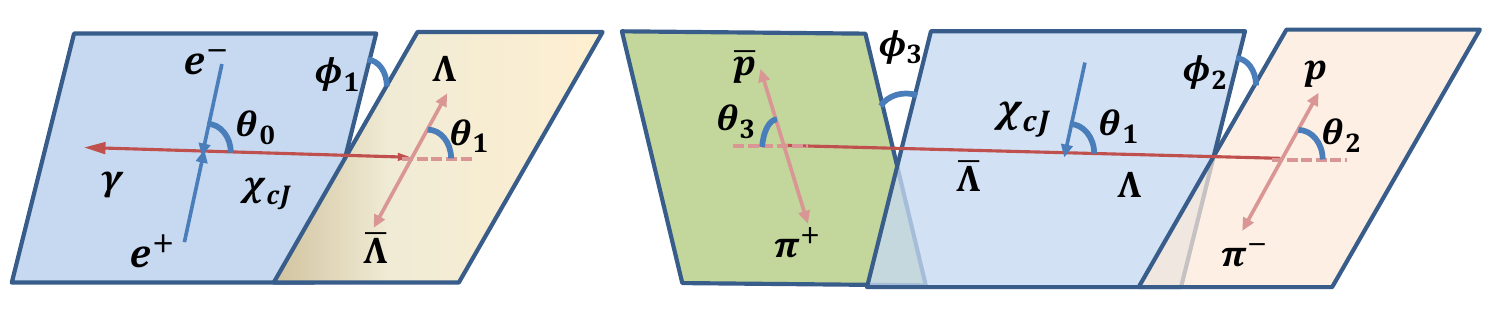}
\caption{Definition of helicity angles for each sequential decays in $e^+e^-\to\psi(3686)\to\gamma\chi_{cJ},~\chi_{cJ}\to\Lambda\bar\Lambda,~\Lambda\to p\pi^-$ and $\bar\Lambda\to\bar p\pi^+$. }
\label{fig:def_helicity_angle}
\end{figure}

The definitions of these amplitudes are summarized in Table~\ref{tab:helicity_def}. The decay topology for each sequential decay is shown in Fig.~\ref{fig:def_helicity_angle}. The procedure for constructing the total amplitude follows the analysis presented in Ref.~\cite{BESIII:2022udq} and is automatically implemented in the software TF-PWA~\cite{jiangyi_tf-pwa}, developed in the TensorFlow framework~\cite{Abadi:2016kic}. TF-PWA includes mass resolution effects for describing the $\chi_{cJ}$ line shape.

\vspace{0.2cm}

\begin{table}[hbp]
    \centering
    \caption{Helicity angles and amplitude definition for the $\chi_{cJ}$ states with $J=0,1,2$ corresponding to $\chi_{c0}$, $\chi_{c1}$, and $\chi_{c2}$, respectively.}
    \begin{tabular}{lcc}
    \hline
    \hline
     Decays    & Angles & Amplitudes \\
    \hline
    $\psi(3686)(M) \rightarrow \gamma(\lambda_1)\chi_{cJ}(\lambda_2)$   &  $\Omega_{0} = (\theta_0,\phi_0)$& $A_{\lambda_1,\lambda_2}^{(J)}$\\
    $\chi_{cJ}  \rightarrow \Lambda(\lambda_3)\bar{\Lambda}(\lambda_4)$   &  $\Omega_{1} = (\theta_1,\phi_1)$ & $B_{\lambda_3,\lambda_4}^{(J)}$\\
    $\Lambda  \rightarrow p(\lambda_5)\pi^-$   &  $\Omega_{2} = (\theta_2,\phi_2)$ & $F_{\lambda_5}$ \\
    $\bar{\Lambda}  \rightarrow \bar{p}(\lambda_6)\pi^+$   &  $\Omega_{3} = (\theta_3,\phi_3)$ & $G_{\lambda_6}$ \\
    \hline
    \hline
    \end{tabular}
    \label{tab:helicity_def}
\end{table}

The amplitude for the sequential decay $\psi(3686)\to
\gamma\Lambda\bar\Lambda$ with $\Lambda(\bar\Lambda)\to p\pi^-(\bar
p\pi^+)$ is the coherent sum of all resonant and non-resonant (NR)
components with spin-parity $J^P$, expressed as: \begin{widetext}
\begin{equation} \begin{aligned}
 \mathcal{A}_k({\bfx},M,\lambda_1,\lambda_5,\lambda_6)
=\sum_{\lambda_2,\lambda_3,\lambda_4}
 D^{1*}_{M,\lambda_1-\lambda_2}(\Omega_0)
D^{J*}_{\lambda_2,\lambda_3-\lambda_4}(\Omega_1)
D^{1/2*}_{\lambda_3,\lambda_5}(\Omega_2)D^{1/2*}_{\lambda_4,\lambda_6}(\Omega_3)
BW_k(m_{\Lambda\bar\Lambda})A_{\lambda_1,\lambda_2}^{(J)}  B_{\lambda_3,\lambda_4}^{(J)}F_{\lambda_5}G_{\lambda_6},
\label{equ:total_amplitude}
\end{aligned}
\end{equation}
\end{widetext}
where
$\bfx=(\Omega_0,\Omega_1,\Omega_2,\Omega_3,m_{\Lambda\bar\Lambda})$;
$D^{J}_{m,n}(\Omega_i)=D^{J}_{m,n}(\phi_i,\theta_i,0)$ is the
Wigner-$D$ matrix. $BW_k$ is Breit-Wigner propagator for $k$-th
resonance, defined as \begin{equation}
    BW(m) = \frac{1}{m_0^2 - m^2 - im_0\Gamma_0} , 
\end{equation}
where $m_0$ is the mass of the resonance, and $\Gamma_0$ is the width
of the resonance.

The helicity amplitude for the two-body decay, e.g., $A_{\lambda_1,\lambda_2}^{(J)}$ and $B_{\lambda_3,\lambda_4}^{(J)}$, involving spin and helicity states $J_0\to J_1(\lambda_1)+J_2(\lambda_2)$, is described within the $L$-$S$ coupling scheme as
\begin{widetext}
 \begin{equation}
    H_{\lambda_1 \lambda_2}   = \sum_{ls} g_{ls}\sqrt{\frac{2l+1}{2J_0+1}} \langle l0,s\delta|J_0\delta\rangle   \langle J_1 \lambda_1, J_2 -\lambda_2|s\delta \rangle  \left(\frac{q}{q_0} \right)^l  B_l^{\prime}(q,q_0,d), \label{s04:ls_coupling}
\end{equation}
\end{widetext}
where $l$ is the orbital angular momentum quantum number, $s$ is the total spin of the two-body system, $\delta=\lambda_1-\lambda_2$ is the helicity difference, $g_{ls}$ is the coupling constant for each partial wave, $q$ is the momentum of the daughter particle in the rest frame of the parent particle, and $q_0$ is calculated at the nominal mass of the parent particle, $d$ is the radius of the barrier factor, and $B_l^{\prime}$ is the reduced Blatt-Weisskopf barrier factor~\cite{BF_VonHippel:1972fg}, which is explicitly expressed as 

\begin{align*}
   B_0^{\prime}(q,q_0,d) &= 1, \\[1em]
   B_1^{\prime}(q,q_0,d) &= \sqrt{\frac{1+(q_0d)^2}{1+(qd)^2}},
\end{align*}
\begin{equation}
   B_2^{\prime}(q,q_0,d) = \sqrt{\frac{9+3(q_0d)^2+(q_0d)^4}{9+3(qd)^2+(qd)^4}},
\end{equation}
where $d$ takes $3.0~\text{GeV}^{-1}$ in the nominal fit. For NR components, we take $BW_k=1$. 

Equation~\ref{s04:ls_coupling} can be utilized to decompose the helicity amplitudes into $L$-$S$ couplings, with an equivalent number of parameters for both hadronic and weak decays. For instance, the two independent helicity amplitudes can be expanded in terms of the $L=1,3$ waves for the $\chi_{c2}\to \Lambda\bar\Lambda$ decay at $q=q_0$, and the barrier factor is canceled: 
\begin{equation}
\begin{aligned}
   B^{(2)}_{+,+} &= \frac{\sqrt{5}}{5} g_{11}  - \frac{\sqrt{30}}{10} g_{31}  , \\  
    B^{(2)}_{+,-} &= \frac{\sqrt{30}}{10}g_{11} + \frac{\sqrt{5}}{5}g_{31}  ,
\end{aligned}
\end{equation}
where +(-) represents the $\Lambda(\bar{\Lambda})$ helicity 1/2($-$1/2).
However, in the radiative decay $\psi(3686)\to\gamma\chi_{cJ}$, the $L$-$S$ expansion results in redundant terms, requiring additional gauge-invariant conditions to constrain the additional degrees of freedom. In the covariant amplitude method, the surplus photon polarizations can be constrained by employing the Coulomb gauge~\cite{Zou:2002ar}. Here, we eliminate the surplus terms as demonstrated in Ref.~\cite{Jing:2023rwz}.

The $\Lambda$ angular distribution in $\chi_{cJ}$ decays can be formulated as~\cite{Chen:2020pia_Pchicj}: 
\begin{equation}\label{}
{dN\over d\cos\theta_1}=\left\{
\begin{array}{lc}
1 & \text{for $\chi_{c0}$,}\\
1 + \alpha_{\chi_{c1}} \cos^2\theta_1 & \text{for $\chi_{c1}$},\\
 1 + \alpha_{\chi_{c2}} \cos^2\theta_1 + \beta_{\chi_{c2}} \cos^4\theta_1& \text{for $\chi_{c2}$},
\end{array}
\right.
\end{equation}
with
\begin{equation}
    \begin{aligned}
        \alpha_{\chi_{c1}} &= \frac{-2 r_1^2 R_1^2+r_1^2+4 R_1^2-2}{2 r_1^2 R_1^2+r_1^2+2} ,\\
        \alpha_{\chi_{c2}} &= \frac{12 r_2^2 R_2^2-6 r_3^2R_2^2-6 r_2^2-12 R_2^2+12}{3 r_3^2 R_2^2+2 r_2^2+2 r_3^2+2 R_2^2},\\
        \beta_{\chi_{c2}}  &= \frac{-12 r_2^2 R_2^2+3 r_3^2 R_2^2+8 r_2^2-2 r_3^2+18 R_2^2-12}{3 r_3^2 R_2^2+2 r_2^2+2 r_3^2+2 R_2^2},\\
    \end{aligned}
    \label{equ:angular_distribution}
\end{equation}
where $r_1 =  |A^{(1)}_{1,1}|/|A^{(1)}_{1,0}|$, $r_2 =  |A^{(2)}_{1,1}|/|A^{(2)}_{1,0}|$, $r_3 = |A^{(2)}_{1,2}|/|A^{(2)}_{1,0}|$, $R_1  =|B^{(1)}_{++}  |/|B^{(1)}_{+-} |$, $R_2 = |B^{(2)}_{++} |/|B^{(2)}_{+-}|$, $R_1 = 0 $ due to the CP conservation in $\chi_{c1}\to\Lambda\bar\Lambda$ decay.

\section{AMPLITUDE FIT PROCESS AND RESULT}\label{sec_fit}

Using the amplitude defined in Eq.~\ref{equ:total_amplitude}, the
total amplitude is the coherent sum of all components, as expressed
by: \begin{equation} |\mathcal{A}(\bfx)|^2 = \frac{1}{2}
\sum_{M=\pm1,\lambda_1,\lambda_5,\lambda_6}\left|\sum_k\mathcal{A}_k(\bfx,M,\lambda_1,\lambda_5,\lambda_6)\right|^2,
\end{equation} where we choose $M=\pm1$ to ensure the helicity
conservation of the $\psi(3686)$ particle generated from the $e^+e^-$
annihilation process. The probability density function is defined as:

\begin{equation}
  P'(\bfx)  = \frac{\left| \mathcal{A}(\bfx)  \right|^2 \epsilon(\bfx) }{ \int \left| \mathcal{A}(\bfx)  \right|^2 \epsilon(\bfx) d\Phi} = \frac{N'(\bfx)}{D}, 
  \label{equ:P_prime}
 \end{equation}
 where $N'(\bfx) = \left| \mathcal{A}(\bfx) \right|^2 \epsilon(\bfx)$ is the numerator, $D = \int \left| \mathcal{A}(\bfx) \right|^2 \epsilon(\bfx) d\Phi$ is the normalization factor (denominator), $\epsilon(\bfx)$ is the detection efficiency, and $\Phi$ is the phase space for $\psi(3686)\to\gamma \Lambda\bar\Lambda$ with $\Lambda(\bar\Lambda)\to p\pi^-(\bar p\pi^+)$. The normalization factor $D$ can be calculated by MC integration over a phase space MC sample.


Considering the mass resolution effects of the observed $\chi_{cJ}$ states in the detected $\Lambda\bar\Lambda$ mass distribution, the numerator in Eq.~\ref{equ:P_prime} changes to:
\begin{equation}
\begin{aligned}
  N(\bfx) & =  \int \left| \mathcal{A}(\bfx_i)   \right|^2  f(m_t|m)  \epsilon(\bfx) dm_t,  \\
  \end{aligned}
\end{equation}
where $m_t$ is the true value of the invariant mass of
$\Lambda\bar{\Lambda}$ and $m$ is the measured
  value. $f(m_t|m)$ represents the probability of the true mass being
  $m_t$ given the measured mass $m$.
The efficiency term
  $\epsilon(\bfx)$ can be discarded because it only contributes a
  constant term to the log-likelihood. By integrating over all
  possible $m_t$, we obtain the probability of observing an event with
  a measured mass of $m$, and $f(m_t| m)$ can be obtained from the
  phase space MC simulation.
The probability density function with mass resolution is defined as:
\begin{equation}
 P(\bfx) = \frac{N(\bfx)}{D}. 
 \label{equ:P}
\end{equation}

The amplitude model parameters are determined by minimizing the negative log-likelihood (NLL). The total NLL is obtained by subtracting the NLL of background contributions from the NLL of the data:

\begin{equation}
    \mathcal{S}=-\ln L = -  \sum_{i\in \text{data} } \ln P(x_i)  + \sum_{i\in \text{bkg} } \ln P(x_i).
\label{equ:NLL}
\end{equation}

The fit fraction (FF) for each component can be calculated as 
\begin{equation}
    FF_i = \frac{\int\left| \mathcal{A}_i  \right|^2 d \Phi}{\int\left|  \mathcal{A} (\bfx) \right|^2 d \Phi}, 
\end{equation}
where $\mathcal{A}_i $ is the amplitude of the $i$-th component and the integration is calculated using the truth-level phase space MC events.

To model the data, we incorporate the three $\chi_{cJ}$ states with their masses and widths fixed to PDG values~\cite{ParticleDataGroup:2024cfk}, except for the $\chi_{c0}$ width, which is treated as a free parameter. Additionally, we include a NR component of $\Lambda\bar\Lambda$ with $J^P=2^+$, determined by the significance test.

\begin{figure*}
    \centering
    \subfigure[]{\includegraphics[width=0.315\textwidth]{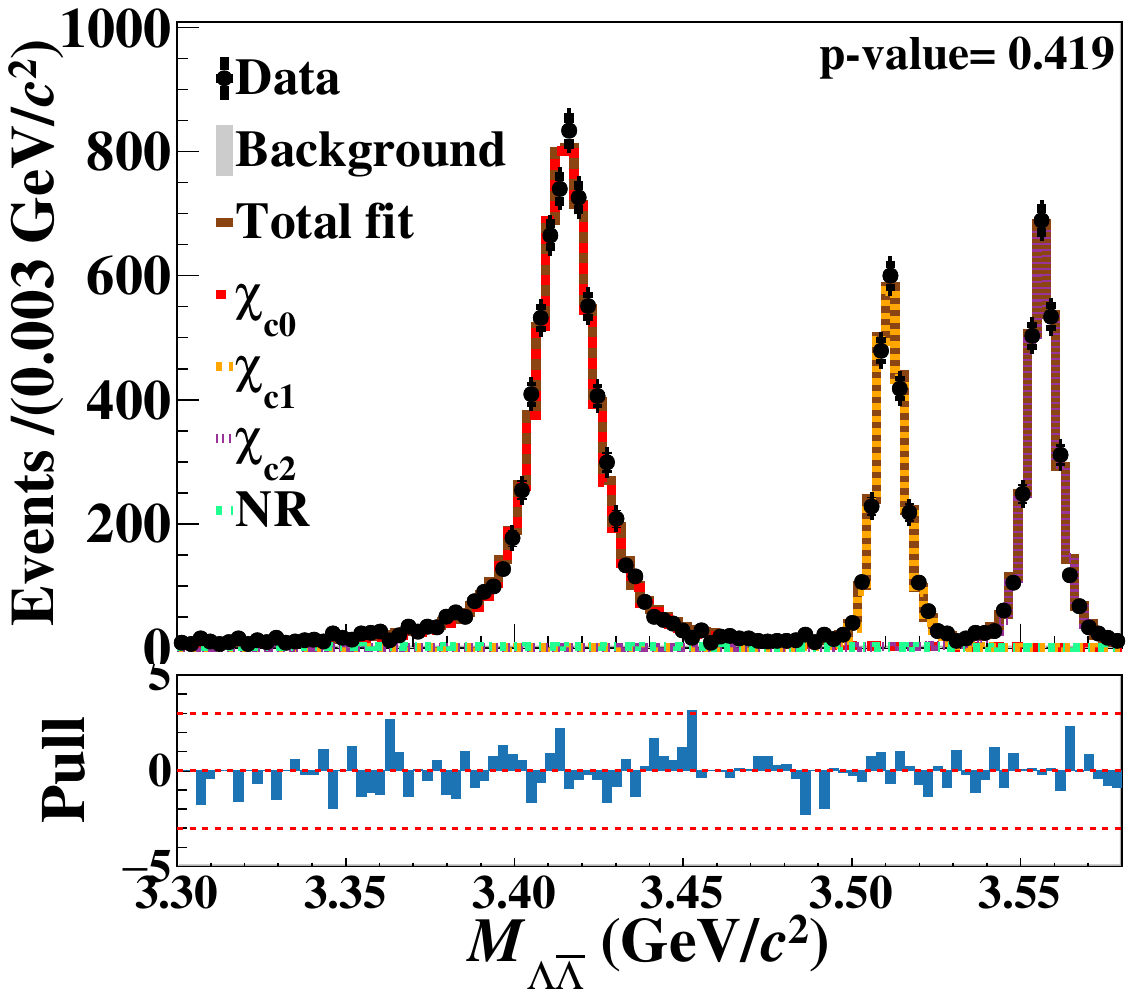}}
    \subfigure[]{\includegraphics[width=0.315\textwidth]{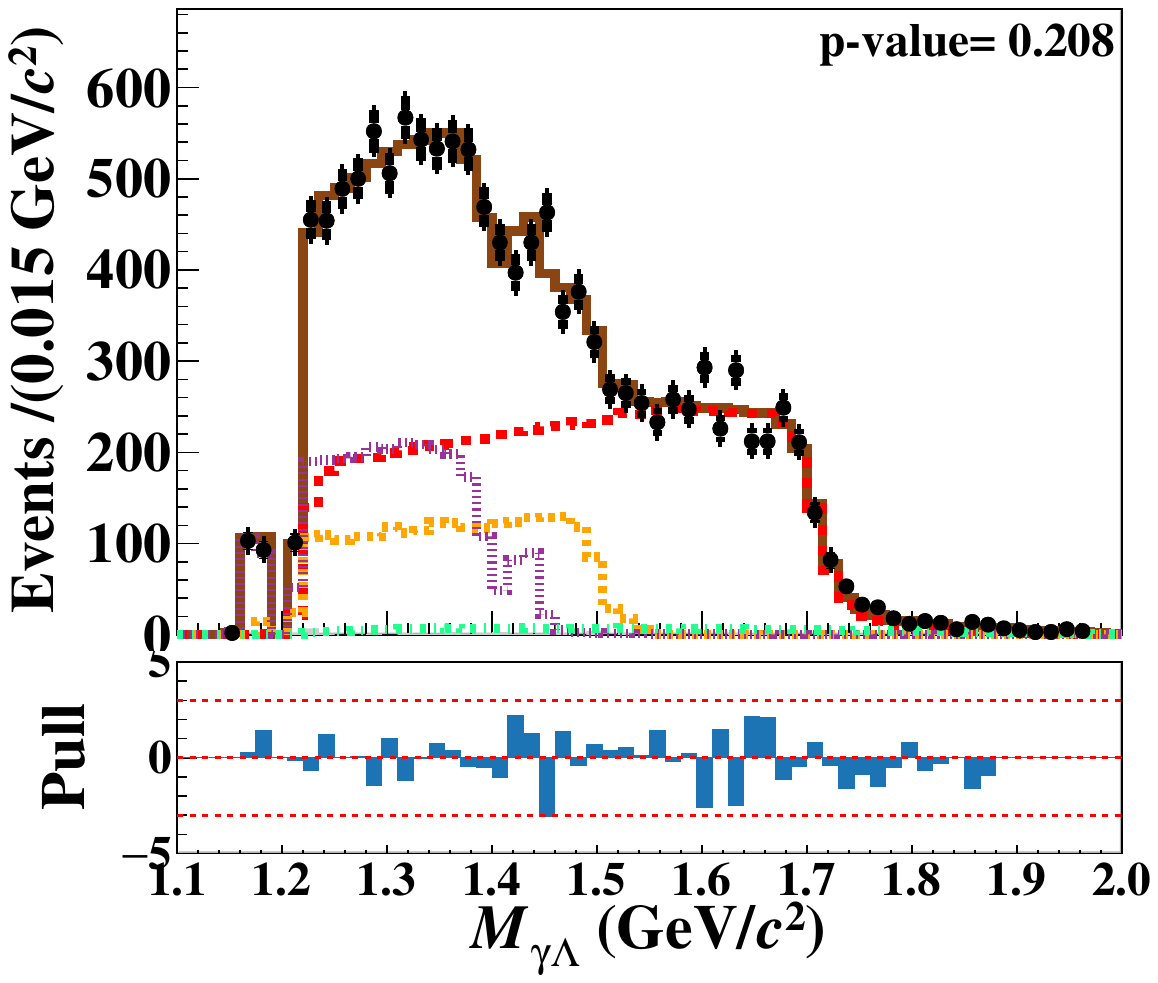}}
    \subfigure[]{\includegraphics[width=0.315\textwidth]{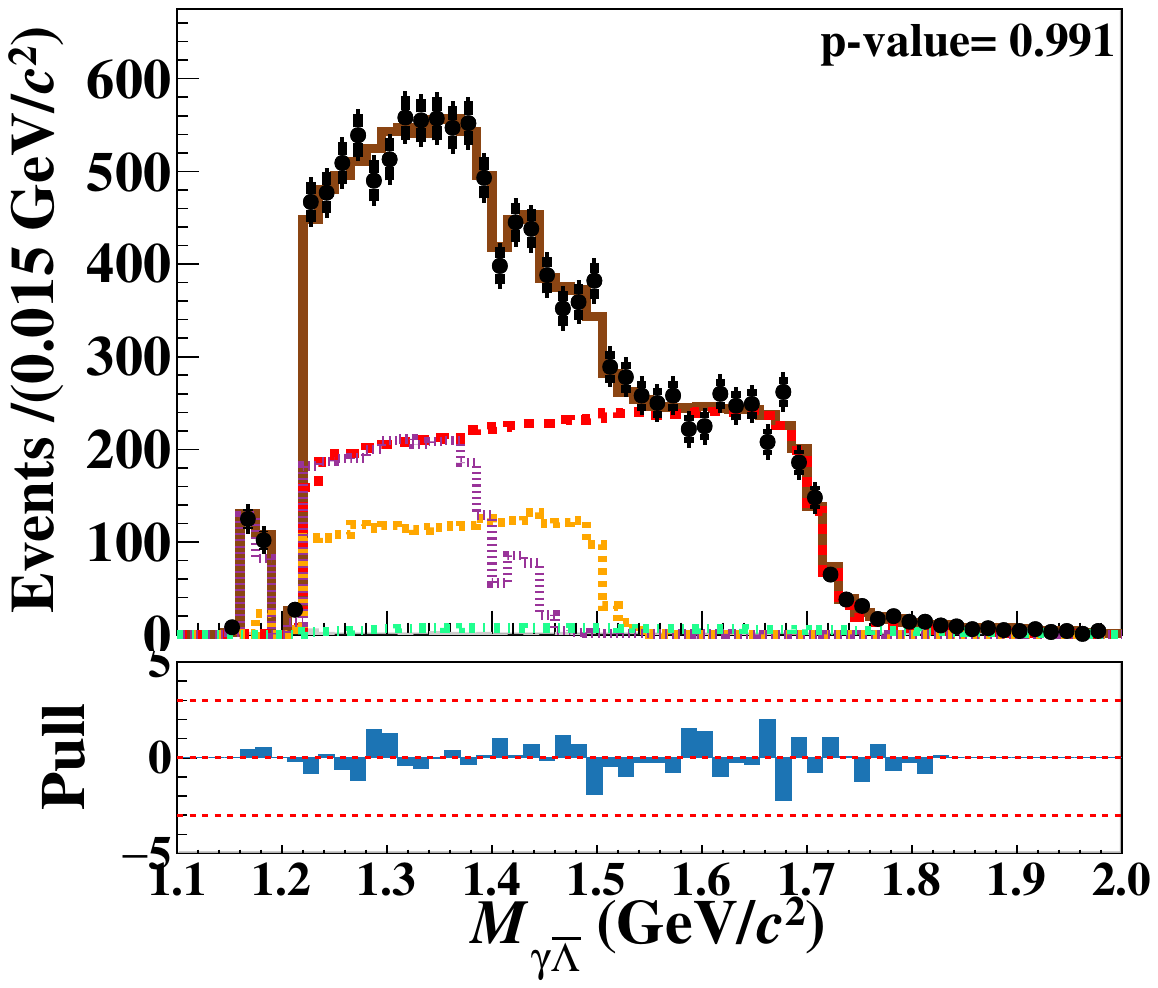}}\\
    \subfigure[]{\includegraphics[width=0.315\textwidth]{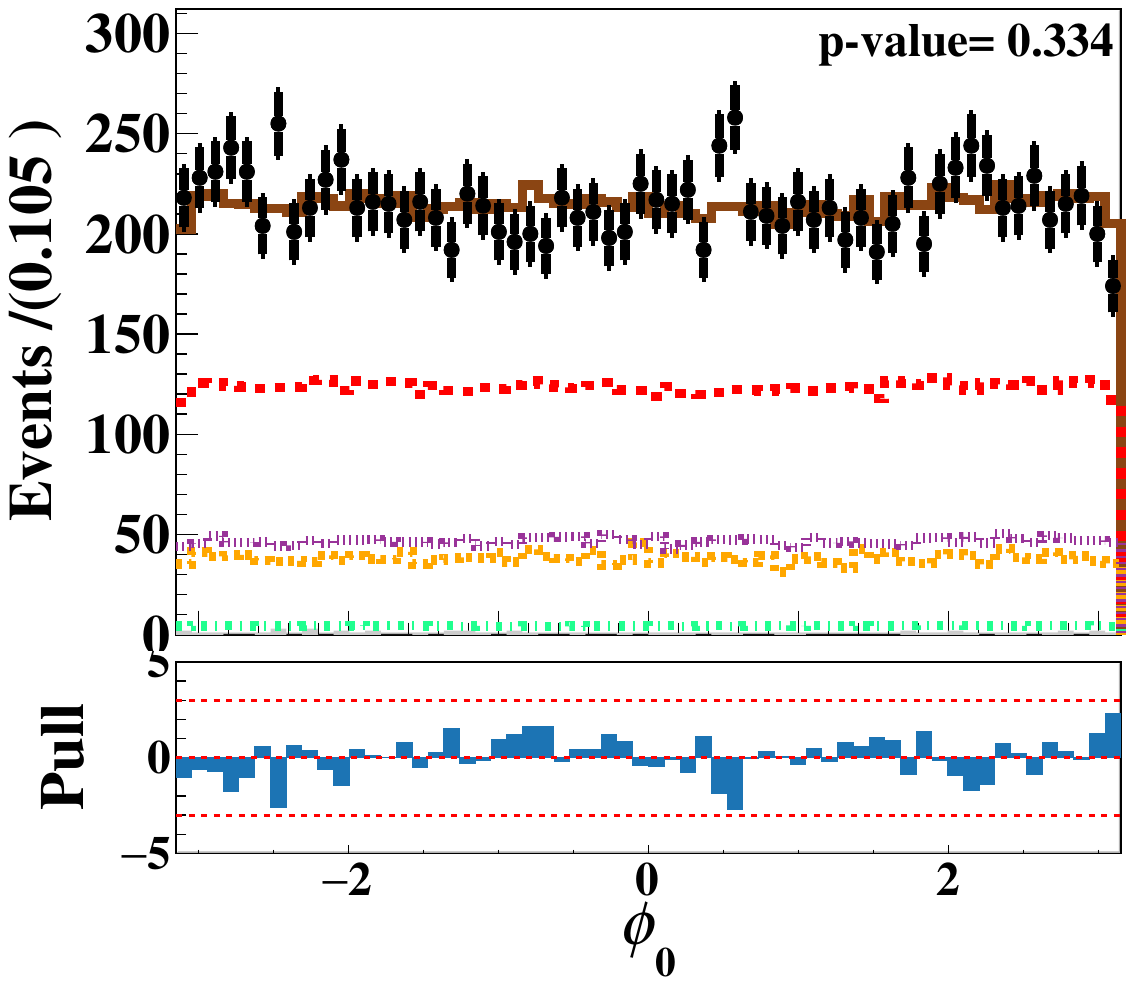}}
    \subfigure[]{\includegraphics[width=0.315\textwidth]{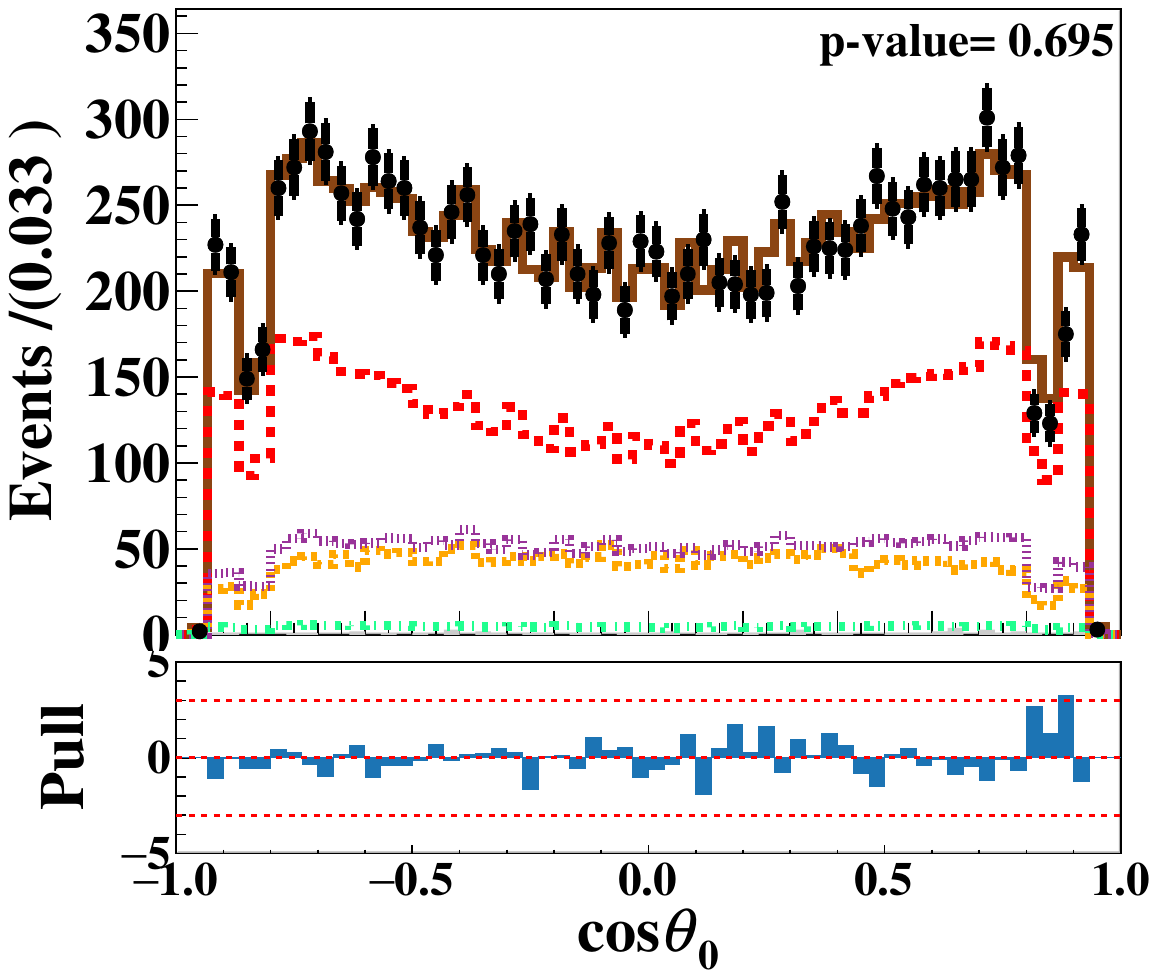}}  
    \subfigure[]{\includegraphics[width=0.315\textwidth]{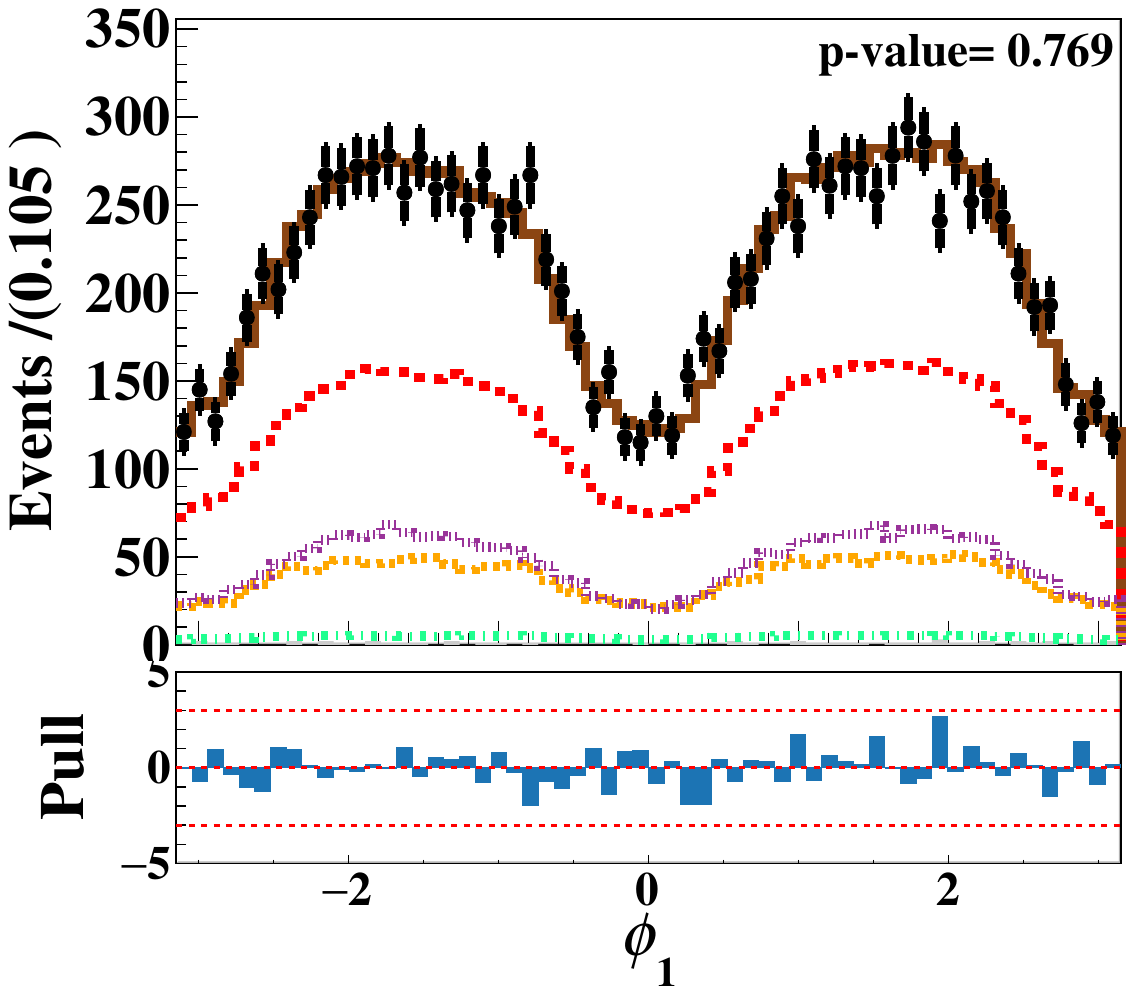}}\\
    \subfigure[]{\includegraphics[width=0.315\textwidth]{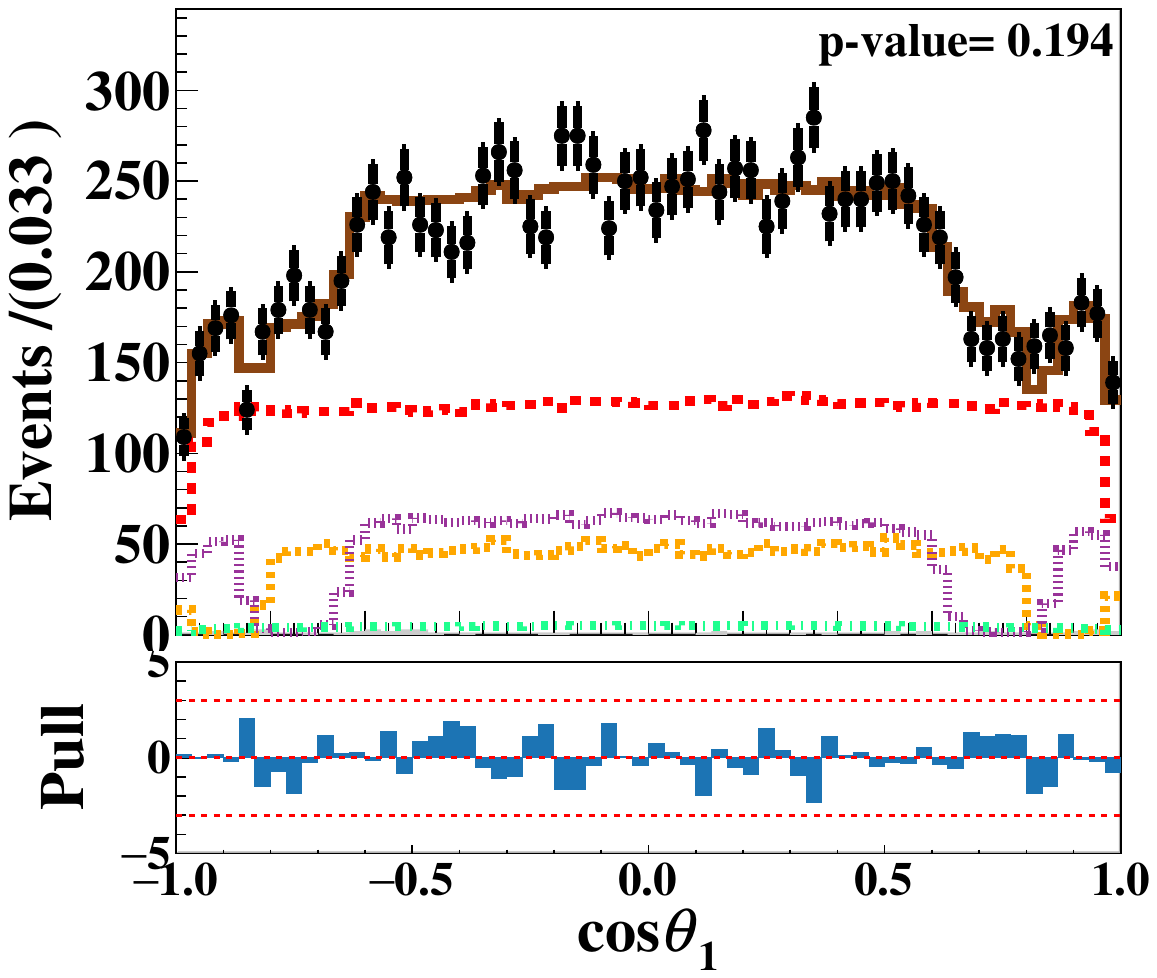}}
    \subfigure[]{\includegraphics[width=0.315\textwidth]{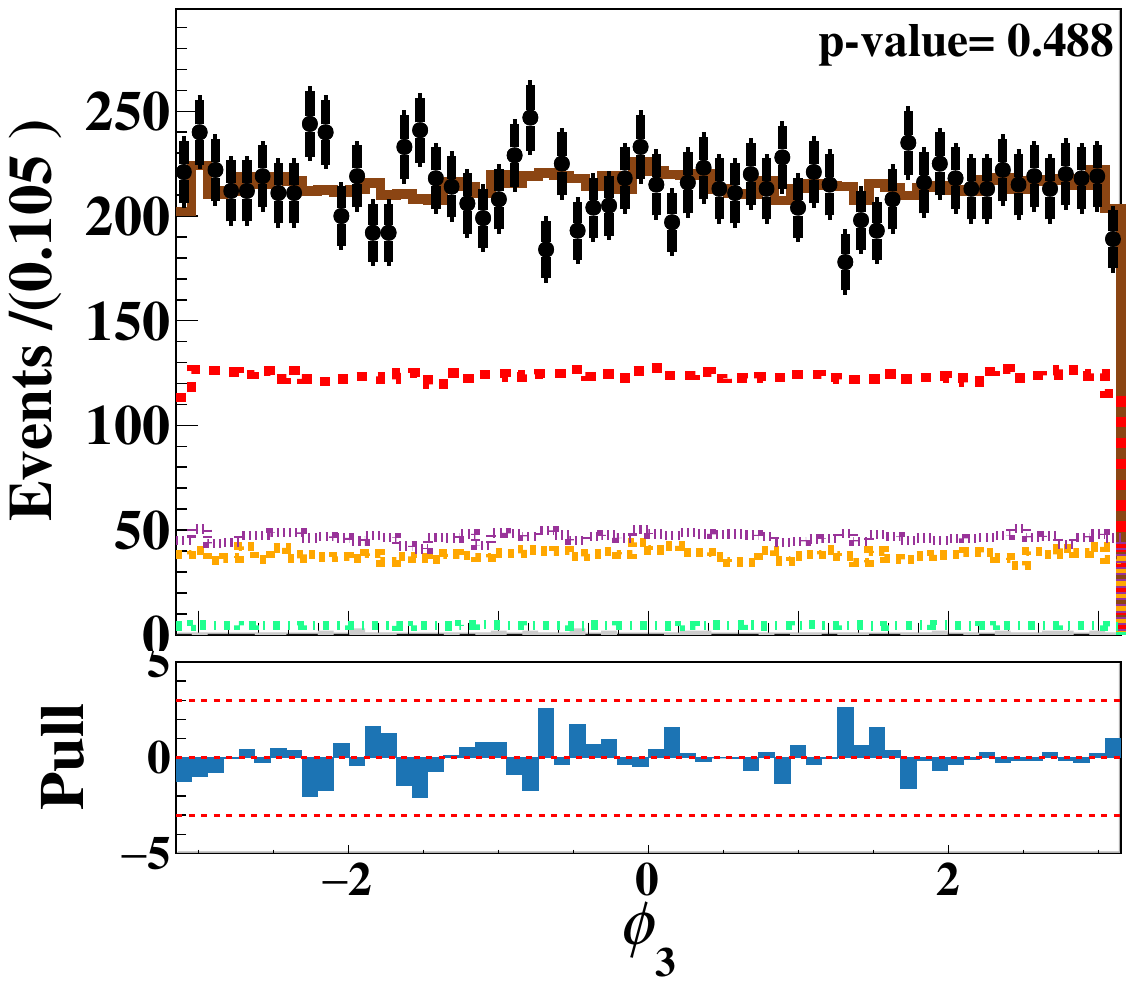}}
    \subfigure[]{\includegraphics[width=0.315\textwidth]{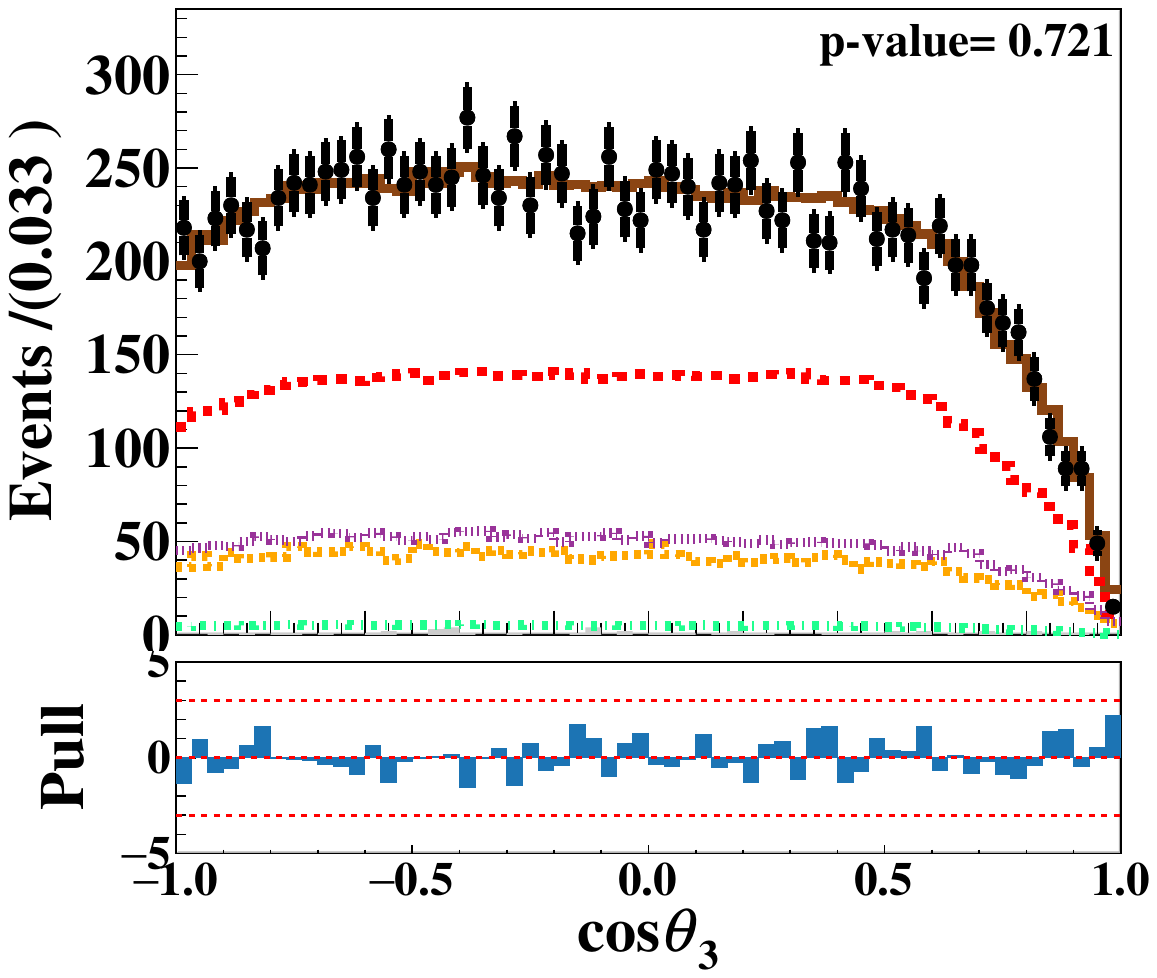}}

\caption{Nominal fit results. (a), (b), (c) are the invariant mass
distribution for $\Lambda\bar{\Lambda}$, $\gamma\Lambda$ and
$\gamma\bar{\Lambda}$, respectively. (d), (e) are the helicity angle
distribution of $\chi_{cJ}$ ($\phi_{0}$,cos$\theta_{0}$), (f), (g)
are the helicity angle distribution for $\Lambda$
($\phi_{1}$,cos$\theta_{1}$), and (h), (i) are the helicity angle
distribution for the proton ($\phi_{3}$,cos$\theta_{3}$). The helicity
angle distribution for anti-proton are similar. In all figures,
the black dots with error bars are data, the brown histograms are the
total fit results, and the color histograms are the line shapes for each
of the components, defined in the legend in figure (a). The gray
histograms are the background contributions. Figure (a) uses 100 bins, 
while the other figures use 60 bins. The $p$-value shown in each figure 
is calculated assuming a $\chi^2$ distribution with $n_{\rm bin}$ degrees 
of freedom, where $n_{\rm bin}$ is the number of bins in each projection. 
This value serves only to assess the consistency between data and model 
in each projection, and is not used in the parameter estimation.
}

    \label{fig:fit_data}
\end{figure*}

The mass and angular projections of the data and fit results are
displayed in Fig.~\ref{fig:fit_data}, where the fit results exhibit good
agreement with data. The helicity amplitude
ratio, fit fractions, angular parameters of $\chi_{c1}$ and
$\chi_{c2}$, and the phase angle of the amplitude are provided in
Table~\ref{tab:fit_result}.

The branching fractions of $\chi_{cJ} \rightarrow \Lambda
\bar{\Lambda}$ are calculated by \begin{equation}
  \mathcal{B}(\chi_{cJ}\to\Lambda \bar{\Lambda}) =
  \frac{N_{\text{obs}} }{\epsilon N_{\psi(3686)}
    \mathcal{B}(\psi(3686)) \mathcal{B}(\Lambda)
    \mathcal{B}(\bar{\Lambda})}\cdot FF_{J}, \end{equation}
where $N_{\text{obs}}=\Nobs$ represents the number of observed signal
events after subtracting the background candidates from the
total number of data events, $\epsilon$ denotes the detection
efficiency, which is determined to be $\Efficiency$\% from a signal MC
sample generated using the amplitude model with fixed parameters based
on the PWA fit. $\mathcal{B}(\psi(3686))$, $\mathcal{B}(\Lambda)=
\BRLambdaPpi$\%, and $\mathcal{B}(\bar{\Lambda})= \BRLambdaPpi$\%, which
are the branching fractions for $\psi(3686)\to\gamma\chi_{cJ}$,
$\Lambda\to p \pi^-$, and $\bar{\Lambda}\to \bar{p}\pi^+$,
respectively, are taken from the PDG~\cite{pdg}.

The number of observed signal events for each $\chi_{cJ}$ state,
measured product branching fractions of $\psi(3686) \rightarrow \gamma
\chi_{cJ}$ and $\chi_{cJ} \rightarrow \Lambda \bar{\Lambda}$ , the
PDG values of $\psi(3686) \rightarrow \gamma \chi_{cJ}$, the
measured values of $\chi_{cJ} \rightarrow \Lambda \bar{\Lambda}$, and
the PDG values of  $\chi_{cJ} \rightarrow \Lambda \bar{\Lambda}$ are
listed in Tab.~\ref{tab:bxxx}. The number of signal events is provided as
a statistical reference and is not used for the branching fraction
calculation.

\begin{table}[htbp]

\caption{Fit results with statistical and systematic
uncertainties where $r_1$, $r_2$, $r_3$, $R_{2}$ are the modulus of amplitude
ratios for $A^{(1)}_{1,1} / A^{(1)}_{1,0}$,
$A^{(2)}_{1,1}/A^{(2)}_{1,0}$, $A^{(2)}_{1,2}/A^{(2)}_{1,0}$ and
$B^{(2)}_{++} /B^{(2)}_{+-}$, $\Delta\Phi_{r_1}$, $\Delta\Phi_{r_2}$,
$\Delta\Phi_{r_3}$, $\Delta\Phi_{R_2}$ are the corresponding relative
phase angles. $FF_{\chi_{c0}}$, $FF_{\chi_{c1}}$, $FF_{\chi_{c2}}$, $FF_{NR(2^+)}$ are
the fit fractions of $\chi_{c0}$, $\chi_{c1}$, $\chi_{c2}$, $NR(2^+)$,
respectively. $\alpha_{\chi_{c1}}$, $\alpha_{\chi_{c2}}$,
$\beta_{\chi_{c2}}$ are the angular parameters of $\chi_{c1}$ and
$\chi_{c2}$ in Eq.~\ref{equ:angular_distribution} and $\Gamma_{\chi_{c0}}$
is the width of $\chi_{c0}$.}
\centering
    \begin{tabular}{lcc}
    \hline
    \hline
    Parameters & Value  & Unit  \\
    \hline 
$r_1$                    & $\rOne$            & -   \\
$r_2$                    & $\rTwo$            & -   \\
$r_3$                    & $\rThree$          & -   \\
$\Delta\Phi_{r_1}$       & $\DeltaPhiROne$    & rad \\
$\Delta\Phi_{r_2}$       & $\DeltaPhiRTwo$    & rad \\
$\Delta\Phi_{r_3}$       & $\DeltaPhiRThree$  & rad \\
$R_{\chi_{c2}}$          & $\RChicTwo$        & -   \\
$\Delta\Phi_{\chi_{c2}}$ & $\DeltaPhiChicTwo$ & rad \\
$FF_{\chi_{c0}}$         & $\FFSJChicZero$      & -   \\
$FF_{\chi_{c1}}$         & $\FFSJChicOne$       & -   \\
$FF_{\chi_{c2}}$         & $\FFSJChicTwo$       & -   \\
$FF_{NR(2^+)}$           & $\FFNRTwoPlus$     & -   \\
$\alpha_{\chi_{c1}}$     & $\AlphaChicOne$    & -   \\
$\alpha_{\chi_{c2}}$     & $\AlphaChicTwo$    & -   \\
$\beta_{\chi_{c2}}$      & $\BetaChicTwo$     & -   \\
$\Gamma_{\chi_{c0}}$     & $\GammaChicZero$   & MeV \\
    \hline
    \hline
    \end{tabular}
    \label{tab:fit_result}
\end{table}

\begin{table*}[t]
\caption{The number of observed signal events for each
$\chi_{cJ}$ state, the measured product branching fractions of $
\psi(3686) \rightarrow \gamma \chi_{cJ}$ and $\chi_{cJ} \rightarrow
\Lambda \bar{\Lambda}$ ($\mathcal{B}_1\mathcal{B}_2$), the PDG branching
fractions of $ \psi(3686) \rightarrow \gamma \chi_{cJ}$ ($\mathcal{B}_1$)
from the PDG~\cite{pdg}, the measured and the PDG~\cite{pdg} branching fractions of
$\chi_{cJ} \rightarrow \Lambda \bar{\Lambda}$ ($\mathcal{B}_2$). The first
uncertainty is statistical and the second is systematic.}
\centering
    \setlength{\tabcolsep}{8pt}
    \begin{tabular}{lccccc}
    \hline
    \hline
    Chan & Events & Meas. $\mathcal{B}_1\mathcal{B}_2$ ($\times 10^{-5}$) & PDG $\mathcal{B}_1$ ($\times 10^{-2}$) & Meas. $\mathcal{B}_2$($\times 10^{-4})$ & PDG $\mathcal{B}_2$($\times 10^{-4}$)  \\
    \hline
$\chi_{c0}$ & $\EventschicZero$ & $\ProdBFChicZero$& $\BRPsipGamchicZero$   & $\BFChicZero$ & $\PDGBFChicZero$    \\
$\chi_{c1}$ & $\EventschicOne$ & $\ProdBFChicOne$  & $\BRPsipGamchicOne$    & $\BFChicOne$  & $\PDGBFChicOne$ \\
$\chi_{c2}$ & $\EventschicTwo$ & $\ProdBFChicTwo$  & $\BRPsipGamchicTwo$    & $\BFChicTwo$  & $\PDGBFChicTwo$  \\
    \hline
    \hline
    \end{tabular}
    \label{tab:bxxx}
\end{table*}

\section{SYSTEMATIC UNCERTAINTY OF THE AMPLITUDE ANALYSIS}\label{sec_sys_amp}

In this section, uncertainties originating from multiple sources are
estimated, encompassing photon detection, $\Lambda(\bar{\Lambda})$
reconstruction, $\Lambda$ decay parameters, Blatt-Weisskopf barrier
factor, additional non-resonant contributions, resonance model,
resolution model, $E_1$ transition, kinematic fit, and non-resonant
model. The results are listed in
Table~\ref{tab:sys_amp}.

\begin{itemize}  

\item  Photon detection and $\Lambda(\bar{\Lambda})$ reconstruction

To estimate the uncertainty due to the data-MC efficiency difference,
we add an additional weight factor to correct the MC efficiency and
perform a new PWA fit. For the photon detection efficiency, the correction
factor is determined using the ISR process 
$J/\psi\to\mu\mu\gamma_{ISR}$ as a control sample. For the
$\Lambda(\bar{\Lambda})$ reconstruction efficiency correction, we use
angular-dependent efficiency correction factors, which are determined
using the control sample
$\psi(3686)\rightarrow\Lambda\bar{\Lambda}$. The differences of the PWA
results between the new fitting and the nominal one are taken as the
systematic uncertainties.

\item  Decay parameter of $\Lambda(\bar{\Lambda})$

In the nominal fit, the decay parameter of $\Lambda(\bar{\Lambda})$ is
fixed to the value measured by BESIII~\cite{BESIII:2022qax}. To account for the systematic
uncertainty, we change the value of decay parameter to 1  standard deviation
larger, and the differences between the
new and nominal results are taken as the systematic
uncertainties.

\item Blatt-Weisskopf barrier factor

In the nominal fit, we incorporate the Blatt-Weisskopf barrier factor into the amplitude model to account for the energy-dependent behavior of the helicity amplitude. To estimate the systematic uncertainty, we remove the Blatt-Weisskopf barrier factor, and the differences between the fitted and nominal values are taken as the systematic uncertainties.

\item Additional non-resonant components

To estimate the systematic uncertainty related to additional
non-resonant components, we add $NR(0^-)$, $NR(0^+)$, and
$NR(1^+)$ one at a time to the base model and perform alternative fits. The largest
differences between the fitted and nominal results are taken
as the systematic uncertainties.

\item Resonance Model

In the nominal fit, we use a constant width Breit-Wigner propagator
for $\chi_{c0}$. To estimate the systematic uncertainty due to the
resonance model, we perform an alternative fit using a running width
Breit-Wigner propagator for $\chi_{c0}$, with the running width
$\Gamma(m) = \Gamma_0(\frac{q}{q_0})^{2l+1}\frac{m_0}{m} B_l^{\prime
2}(q,q_0,d)$. The differences between the fitted and nominal values are taken as the systematic uncertainties.

\item  Resolution model

The data-MC resolution difference is determined by fitting the
$\chi_{c1}$ and $\chi_{c2}$ data using the signal MC shape convolved
with a Gaussian function. The standard deviation of the Gaussian
function is determined to be $(1.47 \pm 0.16)$~MeV.  To
evaluate the systematic uncertainty, we enlarge the data-MC resolution
difference by $1$ standard deviation and get a new resolution model. The
differences between the PWA results using the new resolution model and
the nominal results are taken as the systematic uncertainties.

\item  $E_1$ transition

To estimate the systematic uncertainty due to the $E_1$ transition, we
multiply the Breit-Wigner propagator with the $E_1$ transition factor
$(\frac{E_\gamma}{E^0_\gamma})^{\frac{3}{2}}$ and damping factor
$D(\sqrt{s}) = (\frac{(E_\gamma^0)^2}{E_\gamma^0E_\gamma + (E_\gamma^0
- E_\gamma)^2})^{\frac{1}{2}}$~\cite{Anashin:2010dh} and perform a
new PWA fit. The differences between the new PWA and nominal PWA results are taken as the systematic uncertainties.

\item  Kinematic fit 

In order to take into account the impact of the kinematic fit on the PWA results, we
perform PWA fits using two sets of MC samples: with and without helix parameter correction \cite{BESIII:2012mpj}. The differences between
the two results are taken as the systematic uncertainties.

\item  Non-resonant model

 To evaluate the uncertainty due to the NR model, we performed a new fit with varying the NR propagator from $NR=1.0$ to a first-order
 polynomial, $NR=1.0+ r e^{i \phi} m$, where $r$ and $\phi$ are
 arbitrary parameters and $m$ is the invariant mass of $\Lambda\bar{\Lambda}$. The
 differences between the new PWA and nominal PWA results
 are taken as the systematic uncertainties.
 
\end{itemize}

\begin{table*}
    \caption{The relative systematic uncertainty in the PWA
      results, expressed as a percentage. I is from the photon
      reconstruction and $\Lambda(\bar{\Lambda})$ reconstruction, II
      is from the $\Lambda(\bar{\Lambda})$ decay parameter,  III is
      from using the Blatt-Weisskopf barrier factor in helicity
      amplitude, IV is from the non-resonant components,  V is from
      the resonance model, VI is from the resolution difference
      between data and MC, VII is from the influence of the $E1$
      transition and damping factor, VIII is from the uncertainty of
      kinematic fit, and IX is from the uncertainty of NR model.
      The total systematic uncertainty is calculated as the quadrature sum of the individual uncertainties.}
    \begin{center}
        \setlength{\tabcolsep}{12.5pt}
\begin{tabular}{l|cccccccccccc}
    \hline
    \hline
    Parameter   & I    & II    & III    & IV & V & VI  & VII & VIII & IX & Total     \\
    \hline
$r_1$                      &       0.20  &       0.06  &       2.56  &       1.29  &       0.00  &       0.02  &       0.09  &       0.07  &       0.02  &       2.88  \\  
$r_2$                      &       1.31  &       0.06  &       4.33  &       1.66  &       0.00  &       0.05  &       0.97  &       0.14  &       0.21  &       4.92  \\  
$r_3$                      &       1.03  &       0.05  &       4.81  &       1.95  &       0.01  &       0.06  &       0.98  &       0.05  &       0.32  &       5.40  \\  
$\Delta\Phi_{r_1}$         &       6.26  &       0.10  &      41.26  &      79.69  &       0.39  &       4.67  &      24.97  &       7.52  &       4.09  &      93.86  \\  
$\Delta\Phi_{r_2}$         &       2.77  &       0.02  &      61.49  &      10.25  &       0.03  &       0.56  &      12.11  &       0.38  &       0.12  &      63.57  \\  
$\Delta\Phi_{r_3}$         &       6.87  &       0.38  &      15.21  &      31.05  &       0.41  &       0.13  &      61.62  &       0.55  &       0.85  &      71.00  \\  
$R_{\chi_{c2}}$            &       1.75  &       0.03  &       1.81  &       1.37  &       0.00  &       0.03  &       1.37  &       0.09  &       0.01  &       3.18  \\  
$\Delta\Phi_{\chi_{c2}}$   &       2.07  &       0.25  &      12.70  &       6.52  &       0.05  &       0.63  &       5.11  &       0.31  &       1.00  &      15.36  \\  
$FF_{\chi_{c0}}$           &       0.02  &       0.00  &       0.11  &       1.39  &       0.00  &       0.03  &       0.05  &       0.06  &       0.06  &       1.40  \\  
$FF_{\chi_{c1}}$           &       0.08  &       0.00  &       0.12  &       1.66  &       0.00  &       0.08  &       0.96  &       0.01  &       0.09  &       1.93  \\  
$FF_{\chi_{c2}}$           &       0.23  &       0.01  &       0.36  &       0.87  &       0.01  &       0.05  &       1.63  &       0.15  &       0.17  &       1.91  \\  
$FF_{NR(2^+)}$             &       0.74  &       0.09  &       9.86  &      34.46  &       0.11  &       0.39  &       1.61  &       0.73  &       1.83  &      35.95  \\  
$\alpha_{\chi_{c1}}$       &       0.59  &       0.17  &       7.39  &       3.74  &       0.00  &       0.06  &       0.26  &       0.22  &       0.05  &       8.31  \\  
$\alpha_{\chi_{c2}}$       &       4.68  &       0.27  &      23.34  &       9.15  &       0.03  &       0.29  &       4.16  &       0.41  &       1.58  &      25.89  \\  
$\beta_{\chi_{c2}}$        &      14.84  &       0.76  &      51.00  &      17.52  &       0.02  &       0.47  &       9.56  &       2.35  &       2.22  &      56.84  \\  
$\Gamma_{\chi_{c0}}$       &       0.10  &       0.01  &       0.05  &       0.18  &       0.01  &       0.54  &       0.70  &       0.00  &       0.26  &       0.94  \\ 
    \hline
    \hline
    \end{tabular}
    
\end{center}
\label{tab:sys_amp}
\end{table*}

\section{SYSTEMATIC UNCERTAINTY OF THE BRANCHING FRACTION}\label{sec_sys_br}

In this section, we estimate systematic uncertainties in the
determination of the branching fraction. The sources of uncertainties
taken into account encompass photon detection, the number of
$\psi(3686)$ events, branching fractions for $\Lambda$ decay, fit
fractions, $\Lambda\bar{\Lambda}$ reconstruction, kinematic fit, and
signal model. The results are listed in Table~\ref{tab:sys_br}.

\begin{itemize}

\item Photon reconstruction

The uncertainty in photon detection efficiency is estimated to be
0.2\% per photon, determined using a control sample of the ISR process,
$J/\psi\to\mu\mu\gamma_{ISR}$.

\item Number of $\psi(3686)$ events and branching fraction of $\Lambda$ decay 

The uncertainties of the number of $\psi(3686)$ events~\cite{BESIII:2024lks}, the branching fraction of $\psi(3686)\rightarrow\gamma\chi_{cJ}$~\cite{pdg}, and the branching fraction of $\Lambda\rightarrow p\pi^-(\bar{\Lambda}\rightarrow \bar{p}\pi^+)$~\cite{pdg} are propagated as systematic uncertainties.

\item Fit fraction

The systematic uncertainties in the $FF$s are propagated to the branching fractions as systematic uncertainties.

\item $\Lambda\bar{\Lambda}$ reconstruction

For the $\Lambda(\bar{\Lambda})$ reconstruction efficiency correction, we use angular-dependent efficiency correction factors, which are determined using the control sample $\psi(3686)\rightarrow\Lambda\bar{\Lambda}$. The uncertainty of efficiency propagated from the efficiency correction factor is taken as the systematic uncertainty.

\item Kinematic fit \\

Due to the data-MC differences in helix parameters \cite{BESIII:2012mpj}, kinematic
fits may lead to discrepancies in the efficiency. To account for this
effect, we calculate the efficiency using two MC samples: one directly
using the MC helix parameters, and another using control samples to
correct the MC helix parameters before performing the kinematic
fit. The difference between these two
efficiencies on the branching fractions are assigned as the systematic
uncertainties.

\item Signal Model

The systematic uncertainty due to the signal model is estimated to be 0.2\% by varying the PWA model within the uncertainty of the model  parameters.

\end{itemize}

\begin{table*}

\caption{The relative systematic uncertainty of the
branching fraction, expressed as a percentage. I is from photon reconstruction, II is from the number of $\psi(3686)$ events, III is from the branching fraction of $\Lambda\rightarrow p\pi^-$, IV is from the branching fraction of $\psi(3686)\rightarrow\gamma\chi_{cJ}$, V is from the $FF$s of the PWA, VI is from the $\Lambda(\bar{\Lambda})$ reconstruction, VII is from the kinematic fit, and VIII is from the signal model. The total systematic uncertainty is calculated as the quadrature sum of the individual uncertainties.}

    \begin{center}
        \setlength{\tabcolsep}{15pt}
\begin{tabular}{l|cccccccccc}
    \hline
    \hline
    Decay & I  & II & III & IV & V & VI & VII & VIII & Total\\
    \hline
$\mathcal{B}(\chi_{c0}  \rightarrow \Lambda\bar{\Lambda})$   & 0.20  & 0.53  & 1.56  & 2.26  & 1.40  & 0.29  & 0.41  & 0.20  & 3.18  \\
$\mathcal{B}(\chi_{c1}  \rightarrow \Lambda\bar{\Lambda})$   & 0.20  & 0.53  & 1.56  & 2.77  & 1.93  & 0.29  & 0.41  & 0.20  & 3.80  \\
$\mathcal{B}(\chi_{c2}  \rightarrow \Lambda\bar{\Lambda})$   & 0.20  & 0.53  & 1.56  & 2.45  & 1.92  & 0.29  & 0.41  & 0.20  & 3.57  \\
    \hline
    \hline
    \end{tabular}
\end{center}
\label{tab:sys_br}
\end{table*}

\section{SUMMARY}

Using $\LumiPsip$ million $\psi(3686)$ events collected by the BESIII
detector, we conduct a partial wave analysis of the
decay process $\psi(3686) \rightarrow \gamma\chi_{cJ} \rightarrow
\gamma\Lambda\bar{\Lambda}$. The helicity amplitude ratio for
$\chi_{c2}$ is measured for the first time to be $R_{\chi_{c2}} = \RChicTwo$, with a
relative phase angle of $\Delta\Phi_{\chi_{c2}} = \DeltaPhiChicTwo$~rad. We also report angular distribution parameters for $\chi_{c2}$:
$\alpha_{\chi_{c2}} = \AlphaChicTwo$ and $\beta_{\chi_{c2}} =
\BetaChicTwo$. The width of $\chi_{c0}$ is determined to be
$\GammaChicZero$~MeV, providing a precise measurement of this
parameter. Branching fractions for $\chi_{cJ} \rightarrow
\Lambda\bar{\Lambda}$ $(J = 0, 1, 2)$ are measured to be
$(\BFChicZero) \times 10^{-4}$, $(\BFChicOne) \times 10^{-4}$, and
$(\BFChicTwo) \times 10^{-4}$, respectively. These results demonstrate
a notable enhancement in precision compared to current world average
values~\cite{pdg}.

The BESIII Collaboration thanks the staff of BEPCII (https://cstr.cn/31109.02.BEPC) and the IHEP computing center for their strong support. This work is supported in part by National Key R\&D Program of China under Contracts Nos. 2023YFA1606000, 2023YFA1606704; National Natural Science Foundation of China (NSFC) under Contracts Nos. 12175244, 11635010, 11935015, 11935016, 11935018, 12025502, 12035009, 12035013, 12061131003, 12192260, 12192261, 12192262, 12192263, 12192264, 12192265, 12221005, 12225509, 12235017, 12361141819; Beijing Natural Science Foundation of China (BNSF) under Contract No. IS23014; the Chinese Academy of Sciences (CAS) Large-Scale Scientific Facility Program; CAS under Contract No. YSBR-101; 100 Talents Program of CAS; CAS Project for Young Scientists in Basic Research No. YSBR-117; The Institute of Nuclear and Particle Physics (INPAC) and Shanghai Key Laboratory for Particle Physics and Cosmology; German Research Foundation DFG under Contract No. FOR5327; Istituto Nazionale di Fisica Nucleare, Italy; Knut and Alice Wallenberg Foundation under Contracts Nos. 2021.0174, 2021.0299; Ministry of Development of Turkey under Contract No. DPT2006K-120470; National Research Foundation of Korea under Contract No. NRF-2022R1A2C1092335; National Science and Technology fund of Mongolia; National Science Research and Innovation Fund (NSRF) via the Program Management Unit for Human Resources \& Institutional Development, Research and Innovation of Thailand under Contract No. B50G670107; Polish National Science Centre under Contract No. 2024/53/B/ST2/00975; Swedish Research Council under Contract No. 2019.04595; U. S. Department of Energy under Contract No. DE-FG02-05ER41374


%

\end{document}